\documentclass[preprint]{aastex62}

\usepackage{enumerate}

\graphicspath{{./}{}}

\received{...}
\revised{...}
\accepted{\today}
\submitjournal{ApJ}

\shorttitle{star-gas surface density correlations I}
\shortauthors{R. Pokhrel et al.}

\begin{document}

\title{Star-Gas Surface Density Correlations in Twelve Nearby Molecular Clouds I:\\
Data Collection and Star-Sampled Analysis} 

\correspondingauthor{Riwaj Pokhrel}
\email{Riwaj.Pokhrel@UToledo.Edu, riwajpokhrel@gmail.com}

\author[0000-0002-0557-7349]{Riwaj Pokhrel}
\affil{Ritter Astrophysical Research Center, Department of Physics and Astronomy, University of Toledo, Toledo, OH 43606, USA}
\affil{Department of Astronomy, University of Massachusetts, 710 North Pleasant Street, Amherst, MA 01003, USA}

\author{Robert A. Gutermuth}
\affil{Department of Astronomy, University of Massachusetts, 710 North Pleasant Street, Amherst, MA 01003, USA}

\author{Sarah K. Betti}
\affil{Department of Astronomy, University of Massachusetts, 710 North Pleasant Street, Amherst, MA 01003, USA}

\author{Stella S. R. Offner} 
\affil{Department of Astronomy, The University of Texas at Austin, 2500 Speedway, Austin, TX 78712, USA}

\author{Philip C. Myers} 
\affil{Harvard-Smithsonian Center for Astrophysics, 60 Garden Street, Cambridge, MA 02138, USA}

\author{S. Thomas Megeath} 
\affil{Ritter Astrophysical Research Center, Department of Physics and Astronomy, University of Toledo, Toledo, OH 43606, USA}

\author{Alyssa D. Sokol}
\affil{Department of Astronomy, University of Massachusetts, 710 North Pleasant Street, Amherst, MA 01003, USA}

\author{Babar Ali}
\affil{Space Sciences Institute, 4750 Walnut Street, Suite 205, Boulder, CO, USA}

\author{Lori Allen}
\affil{National Optical Astronomy Observatory, 950 North Cherry Avenue, Tucson, AZ, 85719, USA}

\author{Tom S. Allen}
\affil{Portland State University, 1825 SW Broadway Portland, OR 97207, USA} \affil{Lowell Observatory, 1400 West Mars Hill Road, Flagstaff, AZ 86001, USA}

\author{Michael M. Dunham}
\affil{Department of Physics, State University of New York at Fredonia, 280 Central Ave, Fredonia, NY 14063, USA}

\author{William J. Fischer} 
\affil{Space Telescope Science Institute, Baltimore, MD 21218, USA}

\author{Thomas Henning} 
\affil{Max-Planck-Institute for Astronomy, K\"onigstuhl 17, D-69117 Heidelberg, Germany}

\author{Mark Heyer}
\affil{Department of Astronomy, University of Massachusetts, 710 North Pleasant Street, Amherst, MA 01003, USA}

\author{Joseph L. Hora}
\affil{Harvard-Smithsonian Center for Astrophysics, 60 Garden Street, Cambridge, MA 02138, USA}

\author{Judith L. Pipher}
\affil{Department of Physics and Astronomy, University of Rochester, Rochester, NY 14627, USA}

\author{John J. Tobin}
\affil{National Radio Astronomy Observatory, 520 Edgemont Road, Charlottesville, VA 22903, USA}

\author{Scott J. Wolk}
\affil{Harvard-Smithsonian Center for Astrophysics, 60 Garden Street, Cambridge, MA 02138, USA}





\begin{abstract}

We explore the relation between the stellar mass surface density and the mass surface density of molecular hydrogen gas in twelve nearby molecular clouds that are located at $<$1.5 kpc distance. The sample clouds span an order of magnitude range in mass, size, and star formation rates. We use thermal dust emission from $Herschel$ maps to probe the gas surface density and the young stellar objects from the most recent $Spitzer$ Extended Solar Neighborhood Archive (SESNA) catalog to probe the stellar surface density. Using a star-sampled nearest neighbor technique to probe the star-gas surface density correlations at the scale of a few parsecs, we find that the stellar mass surface density varies as a power-law of the gas mass surface density, with a power-law index of $\sim$2 in all the clouds. The consistent power-law index implies that star formation efficiency is directly correlated with gas column density, and no gas column density threshold for star formation is observed. We compare the observed correlations with the predictions from an analytical model of thermal fragmentation, and with the synthetic observations of a recent hydrodynamic simulation of a turbulent star-forming molecular cloud. We find that the observed correlations are consistent for some clouds with the thermal fragmentation model and can be reproduced using the hydrodynamic simulations.

\end{abstract}


\section{Introduction} 
\label{sec:intro_sg}

The physical processes that govern the conversion of interstellar gas to stars have been long investigated in both galactic (e.g., \citealt{Schmidt59,Schmidt63,Lada10,Lada12,Evans14}) and extragalactic (e.g., \citealt{Sanduleak,Hamajima75,Kennicutt98,Wong02,Gao04,Suzuki10}) contexts. \cite{Schmidt59,Schmidt63} derived one of the first relations between star formation rate and properties of interstellar gas using distributions of local HI gas and stars orthogonal to the Galactic plane. Popularly known as the ``Schmidt Law," the relation states that the star formation rate density is proportional to the square of the density of the gas. Similar super-linear power-laws were subsequently reported for nearby galaxies (e.g., \citealt{Sanduleak, Hamajima75}). Later works were further expanded to include molecular hydrogen and larger samples of galaxies (e.g., \citealt{Kennicutt89}); and the Schmidt law was further generalized in terms of the \emph{surface densities} of star formation rate (SFR) and gas mass as follows:

\begin{equation} \label{eq:kslaw}
    \Sigma_{\rm{SFR}} \propto \Sigma_{\rm{gas}}^{N}
\end{equation}

\noindent
\cite{Kennicutt98} compiled galaxy averaged measurements of $\Sigma_{\rm{SFR}}$ and $\Sigma_{\rm{gas}}$ using normal spirals and starburst galaxies and found $N$ = 1.4 $\pm$ 0.15. This formulation is widely known as the ``Kennicutt-Schmidt relation". 
Apart from the galactic disk-averaged studies, other efforts to constrain the power-law index $N$ are concentrated on radial or point-by-point measurements on sub-kpc scales (e.g., \citealt{Kuno95, Zhang01, Wong02, Heyer04, Schuster07}), and they reported values of $N$ between 1 and 2. On the spatially resolved 0.1-2 kpc scales, $N$ was found to be around 0.8-1.6 (e.g., \citealt{Kennicutt07,Bigiel08,Braun09,Verley10}). 
\cite{Bigiel08} found a linear relation between $\Sigma_{\rm{SFR}}$ and the molecular hydrogen surface density over the range of 3-50 M$_{\odot}$pc$^{-2}$.
\cite{Gao04} used the dense gas tracer HCN to trace cold molecular gas and total far-IR luminosity for star formation rate. They found a linear correlation between star formation rate and the mass of dense molecular gas (not in terms of their surface densities). Similarly, variations in $N$ have been reported for even smaller scales of 100-500 pc (see reviews by \citealt{Kennicutt12} for details). The variations in $N$ can be attributed to the systematics in data and analysis, such as spatial resolution, fitting techniques, SFR tracer, gas tracer, uncertainties in conversion quantities, etc. Because of these inconsistencies, the underlying physics for a power-law dependence is still an open question.

The extragalactic measurements give star-gas scaling relations based on integrated star formation and gas across multiple clouds. To have a better understanding of the physics responsible for the star-gas scaling relations, scaling laws such as Equation \ref{eq:kslaw} need to be explored at pc/sub-pc scales for clouds in the Milky Way Galaxy using more direct method of measuring the star formation rate rather than high-mass star formation tracers used in extragalactic methods. Such methods use the detected YSOs themselves to infer the star formation rate. \cite{Evans09} found that the SFR for local clouds lie about a factor of $\sim$20 above extragalactic Kennicutt-Schmidt relation and slightly above the extrapolated relation from a study of massive dense clumps \citep{Wu05}. The study by \cite{Evans09} was further extended by \cite{Heiderman10} and they found similar results. Later,
 \cite{Lada10} suggested a surface density threshold of 116 $\pm$ 28 M$_{\odot}$pc$^{-2}$ ($\sim$8 A$_{\rm{V}}$) above which the SFR varies linearly with the mass of dense gas. Similarly, \cite{Heiderman10} reported a steep increase in $\Sigma_{\rm{SFR}}$ with increasing $\Sigma_{\rm{gas}}$ up to $\sim$ 130 M$_{\odot}$pc$^{-2}$, above which they reported a linear scaling relation. \cite{Lada12} further showed that the dispersion between SFR and gas mass was minimized by only including the cloud mass above a threshold of 8 A$_{\rm{V}}$. This picture of low dispersion above 8 A$_{\rm{V}}$ is further supported by \cite{Evans14}.

 On the other hand, \citet{Gutermuth11} found that $\Sigma_{\rm{SFR}}$ $\propto$ $\Sigma_{\rm{gas}}^{2}$ up to several 100 M$_{\odot}$pc$^{-2}$. The square dependence has also been reported by \citet{Lada13} for Orion-A, Taurus, and California giant molecular clouds (GMCs) but with the power-law index of $\sim$3.3 for Orion B. \citet{Gutermuth11} did not find any evidence of a column density threshold for star formation. \citet{Burkert13} argued that the correlations reported by \citet{Lada10} and \citet{Heiderman10} do not require any density threshold but is a consequence of increasing gravitational influence with increasing density. \citet{Clark14} found that the clouds in their simulation still form stars at cloud averaged surface densities that are lower than $\sim$7 A$_{\rm{V}}$. They further suggest that the reports on the threshold of star formation are more likely a consequence of the star formation process, rather than a pre-requisite for star formation. In a study of the Ophiuchus cloud, \citet{Johnstone04} found that cores do not form below 7 A$_{\rm{V}}$ and suggested it to be the threshold extinction for the formation of stars. However, \citet{Sokol19} found a substantial number of sub-mm cores below 7 A$_{\rm{V}}$ in Mon R2 GMC, with no distinct extinction threshold for star formation (also see \citealt{Benedettini18} for $Herschel$ results in Lupus Cloud complex).

This study aims to improve the quality of star-gas surface density correlation constraints derived from nearby clouds. The study is superior to any other such previous works due to the following three reasons. First, to minimize bias in our characterization of the nature of star-gas surface density correlation, our sample consists of twelve molecular clouds with more than an order of magnitude range in mass, size, and star formation rates (see Table \ref{tab:cloudinfo_sg} and \ref{tab:cloudinfo_sg_fit}). Second, we use uniformly reduced datasets for young stellar objects (YSO) and gas. We probe molecular gas using $Herschel$ dust emission maps similar to \citealt{Lombardi14} (N = 1.99 $\pm$ 0.05 for Orion-A, N = 2.16 $\pm$ 0.10 for Orion-B), \citealt{Zari16} (N = 2.4 $\pm$ 0.6 for Perseus), and \citealt{Lada17} (N = 3.31 $\pm$ 0.23 for the California GMC). For the YSOs, we use a new, uniformly reduced catalog from SESNA ($Spitzer$ Extended Solar Neighborhood Archive, Gutermuth et al. in prep.). Third, we use three measurement techniques to explore star-gas density relations that span the size scale from sub-pc to entire clouds. These methods include the star-sampled nearest neighbor technique (e.g., \citealt{Gutermuth11, Masiunas12, Rapson14, Li19}), gas-sampled integrated A$_{\rm{V}}$ contours (e.g., \citealt{Lada12,Evans14}), and gas-sampled differential A$_{\rm{V}}$ contours (e.g., \citealt{Heiderman10}). Due to the large content of this study, we present and discuss only the nearest neighbor technique in this paper. We will include the results of the gas-sampled methods in an upcoming paper.

In \S \ref{sec:obs_sg} we explain our observations and data reduction methods for obtaining H$_2$ gas column density maps and YSOs catalog. In \S \ref{sec:nnd_sg}, we implement the nearest neighbor technique to study star-gas surface density correlations and present its results. The fragmentation processes that can potentially contribute to the observed correlations are discussed in \S \ref{sec:discussion_sg} and finally a brief conclusion of the study is presented in \S \ref{conclusion_sg}.

\section{Observations and data reduction} \label{sec:obs_sg}

We require two kinds of observation to explore the star-gas surface density correlations: one to probe the gas column density distribution in the molecular clouds and another to probe the YSOs that directly trace star formation rate surface density at a wide range of spatial scales over $\sim$0.5 and $\sim$2.5 Myr time scales. We used observations from $Herschel$ to probe the gas content in clouds, and observations from 2MASS,  $Spitzer$ and UKIDSS to probe and classify the stellar sources.

\subsection{N(H$_2$) maps using $Herschel$}
\label{subsec:herschel_sg}

In the past, star-gas surface density correlations have generally been studied using near-IR extinction maps to characterize the spatial distribution of gas column density \citep{Lada10, Gutermuth11, Harvey13, Lada13}. Although these maps have the advantage over molecular line maps \citep{Goodman09}, they have several disadvantages. The angular resolution of such a map depends on the density of background stars that are detected. Also, for dense foreground clouds with A$_{\rm{V}} \gtrsim$ 20 mag, faint background stars are often extinguished beyond detection limits. Consequently, this method does not reliably estimate the dust column density for high extinction regions without deeper photometry. Thus, this method can effectively saturate toward high-density regions. In contrast, {\it Herschel}’s unprecedented angular resolution and sensitivity in the far-IR where the dust itself is emitting strongly enables dust emission maps of superb quality over large areas of sky (c.f., \citealt{Andre10}). To ensure uniformity in gas column density for all the clouds, we use the gas maps derived from $Herschel$ observations only. $Herschel$ OBSIDs for all the clouds utilized in this study are listed in Appendix \ref{app:herschel}.

For the Gould Belt clouds that are $<$500 pc away, we obtained the gas column density maps from the Herschel Gould Belt Survey (HGBS) archive \footnote{http://www.herschel.fr/cea/gouldbelt/en/Phocea/Vie\_des\_labos/Ast/ast\_visu.php?id\_ast=66}, including Ophiuchus, Perseus, Orion-B, Aquila North and Aquila South. The only exception is the column density and temperature map for Orion-A, which we obtained from \citet{Stutz15}. For the clouds located beyond the Gould Belt ($>$500 pc), viz. NGC 2264, S140, AFGL 490, Cep OB3, Mon R2, and Cygnus-X, their far-IR emission was mapped with parallel scan-map mode with the ESA $Herschel$ space observatory \citep{Pilbratt10} using both the Photoconductor Array Camera and Spectrometer, PACS, \citep{Poglitsch10} and the Spectral and Photometric Imaging  REceiver, SPIRE, \citep{Griffin10}. For Mon R2, Cep OB3 and Cygnus-X, we have reduced the Level 1 $Herschel$ observations to obtain the final flux calibrated dust emission maps using $Herschel$ Interactive Processing Environment (HIPE; \citealt{Ott10}). For S140, NGC 2264 and AFGL 490, we used level 2.5 and level 3 reduced $Herschel$ observations from the ESA $Herschel$ Science Archive (HSA\footnote{http://archives.esac.esa.int/hsa/whsa/}). We followed \cite{Pokhrel16} to derive column density and temperature maps from the observations. The process is explained briefly from \S\ref{obs:prelim_sg} to \S\ref{obs:greybody_sg}.

Table \ref{tab:cloudinfo_sg} provides the references for individual clouds. The central coordinates, size and mass above 1 A$_{\rm{V}}$ that are listed in Table \ref{tab:cloudinfo_sg} are calculated using the column density maps that we used for this study. For distance to the cloud, recent $Gaia$ results \citep{Gaia16,Gaia18} are used where available. For the clouds where $Gaia$ distance is not available, we use the distances obtained using maser parallaxes and stellar photometry. The references to the column density maps and distance to the clouds are provided. 

For the Mon R2 GMC, \citet{Kuhn19} used $Gaia$ DR2 to estimate $d$ = 948 pc to the central Mon R2 cluster using kinematics in the young star cluster. Similarly, \citet{Zucker19} provided $d$ = 788 pc for GGD 12-15, a young cluster in the Mon R2 GMC (see their Table 2). The average of these $Gaia$ derived distances is 868 pc. We use the parallax measurements for SESNA sources that are detected in $Gaia$ ($>$300 sources) to estimate the distance to the Mon R2 GMC following the recipe in \citet{Lindegren18}. Our measurements show $d$ = 860 $\pm$ 31 pc, similar to the average of \citet{Kuhn19} and \citet{Zucker19}. Our estimation is not far off from other estimates such as 893$^{+44}_{-40}$ pc by \cite{Dzib16} using VLBI techniques and the historically popular distance of 830 pc \citep{Racine68}. Thus, for our analysis we use $Gaia$ derived distance of 860 pc.

The Cygnus-X star-forming complex is the most distant (d = 1400 pc), most massive ($\sim$1.8 $\times$ 10$^6$ M$_{\odot}$ above 1 $A_{\rm{V}}$) and covers the biggest area ($\sim$ 140 pc $\times$ 160 pc) in our sample of clouds. The huge area imposes the possibility of different distances to different sections of the cloud. However, line observations have shown a relatively consistent distance to the overall star-forming complex \citep{Schneider06, Schneider07, Rygl12}. Despite its large area, slight variances in the regional distance in Cygnus-X cause a negligible effect in our analysis because of its large distance.

\begin{deluxetable}{lcccccc} 
\tablecaption{General properties of clouds \label{tab:cloudinfo_sg}}
\tablehead{\colhead{Clouds} & \colhead{R.A., Dec.} & \colhead{Angular size\tablenotemark{a}} & \colhead{Physical size\tablenotemark{a}} & \colhead{Mass\tablenotemark{b}} & \colhead{Distance} & \colhead{Resolution\tablenotemark{c}}\\ \colhead{ } & \colhead{$\mathrm{J(2000)}$} & \colhead{(deg. $\times$ deg.)} & \colhead{(pc $\times$ pc)} & \colhead{(M$_{\odot}$)} & \colhead{$\mathrm{(pc)}$} & \colhead{(pc)}}
\startdata
Ophiuchus$^{1,2}$ & 16h27m31.2s -24d12m39.6s & 4.8 $\times$ 5.0 & 11.5 $\times$ 12.0 & 3000 & 137$^{8}$ & 0.02 \\
Perseus$^{1,3,4}$ & 03h35m40.8s +31d31m55.2s & 6.2 $\times$ 5.0 & 31.6 $\times$ 25.3 & 6000 & 294$^{9}$ & 0.05 \\
Orion-A$^{5}$ & 05h39m24s -07d18m21.6s & 6.8 $\times$ 8.4 & 49.3 $\times$ 60.9 & 55000 & 418$^{10}$ & 0.07 \\
Orion-B$^{1,6}$ & 05h47m36.96s +00d06m00s & 6.8 $\times$ 8.6 & 49.2 $\times$ 62.4 & 17000 & 418$^{10}$ & 0.07 \\
Aquila North$^{1,3}$ & 18h34m59.04s +00d00m00s & 6.2 $\times$ 5.0 & 46.9 $\times$ 37.5 & 34000 & 436$^{8}$ & 0.08 \\
Aquila South$^{1,7}$ & 18h29m42.72s -02d46m48s & 4.4 $\times$ 4.7 & 32.7 $\times$ 35.4 & 50000 & 436$^{8}$ & 0.08 \\
NGC 2264 & 06h41m07.92s +10d01m33.6s & 1.9 $\times$ 3.2 & 24.4 $\times$ 40.7 & 19000 & 738$^{11}$ & 0.13 \\
S140 & 22h21m15.84s +63d42m46.8s & 1.3 $\times$ 1.3 & 17.0 $\times$ 18.0 & 5000 & 764$^{12}$ & 0.13 \\
AFGL 490 & 03h08m07.44s +59d31m04.8s & 1.5 $\times$ 1.5 & 21.0 $\times$ 21.0 & 16000 & 800$^{13}$ & 0.14 \\
Cep OB3 & 22h56m15.36s +62d10m55.2s & 4.9 $\times$ 3.3 & 69.5 $\times$ 47.5 & 79000 & 820$^{14}$ & 0.14 \\
Mon R2 & 06h08m46.8s -06d23m13.2s & 4.3 $\times$ 4.4 & 62.4 $\times$ 63.6 & 33000 & 860$^{15}$ & 0.14 \\
Cygnus-X & 20h28m34.8s +39d31m37.2s & 5.8 $\times$ 6.7 & 142.2 $\times$ 163.7 & 1796000 & 1400$^{16}$ & 0.24
\enddata
\tablecomments{$^{\rm{(a)}}$The cloud size refers to the $Herschel$ spatial coverage that we utilize in this study. $^{\rm{(b)}}$Cloud mass above 1 A$_{\rm{V}}$. $^{\rm{(c)}}$Spatial resolution corresponding to the angular resolution of SPIRE 500 $\micron$ map of 36$''$.}
\tablerefs{(1) \citealt{Andre10}; (2) Ladjelate et al. in prep; (3) Pezzuto et al. in prep; (4) \citealt{Mercimek17}; (5) \citealt{Stutz15}; (6) Konyves et al. in prep; (7) \citealt{Konyves15}; (8) \citealt{Ortiz18}; (9) \citealt{Zucker19}; (10) \citealt{Yan19}; (11) \citealt{Kuhn19}; (12) \citealt{Hirota08a}; (13) \citealt{Obonyo19}; (14) \citealt{Kun08}; (15) average of \cite{Kuhn19} and \cite{Zucker19}, see \S \ref{subsec:herschel_sg}; (16) \citealt{Rygl12}.
}
\end{deluxetable}

\subsubsection{Primary reduction with HIPE} \label{obs:prelim_sg}

In \cite{Pokhrel16}, we used $Herschel$-SPIRE 250, 350 and 500 $\micron$ dust emission maps to obtain the column density and temperature maps for the Mon R2 GMC. As a synopsis of the work of \citet{Pokhrel16}, first, we matched the resolution of all the maps to the poorest resolution, i.e., 36$''$. Then we used a flux ratio plot, specifically F(350)/F(500) versus F(250)/F(350), to constrain the emissivity index ($\beta$). With this constant $\beta$ assumption we performed a modified black body fit on a pixel-by-pixel basis to obtain column density and temperature maps. We used flux uncertainties, flux ratio plot and variation of temperature with column density to set a limit on temperature above which the pixels are generally consistent with being Rayleigh-Jeans limited and masked them out. The pixels that are saturated due to bright emission at shorter wavelengths are also excluded from further analysis. The excluded pixels are very low in number (e.g., $<$0.5 \% for Mon R2 from \citealt{Pokhrel16}) and are mostly high-temperature pixels so the amount of mass lost in such pixels is also quite low. Finally, we compared the column densities obtained by the $Herschel$ and near-IR extinction map to verify their consistency. We followed the same procedure for NGC 2264, S140, AFGL 490 and Cep OB3. The procedure is a bit different for Cygnus-X; hence, below we explain the data reduction procedure for Cygnus-X.

\subsubsection{Cygnus-X data reduction} \label{cygx_reduction}

The expansive area of Cygnus-X required reducing a large number of observations (see Appendix \ref{app:herschel} for the details of each observation). We used HIPE, version 15.0.1 to reduce PACS and SPIRE observations and finally mosaic them. The raw data were obtained in both in-scan and cross-scan (orthogonal) mode to help mitigate scanning artifacts. We adjusted the standard pipeline scripts to construct combined maps recovering the extended emission from the two sets of scans. All three SPIRE maps are absolute calibrated using the $Planck$-HFI emission, followed by applying relative gains, de-striping in each band and applying the zero-point correction using the standard HIPE technique (see \citealt{Pokhrel16} for details). We follow \cite{Lombardi14} to reduce PACS observations, with the exception that we use ``JScanam", the HIPE implementation of the Scanamorphos algorithm that removes the low-frequency noise in bolometer arrays while recovering extended emission by using the ``galactic" option \citep{Gracia17}. 

The PACS data products do not include emission that is extended on scales comparable to the map size. While this is not a problem for doing point source photometry in individual maps, it poses a serious problem for obtaining gas column density maps. We follow \cite{Lombardi14} to calibrate the PACS observation and also to do a sanity check based on the calibration of the Planck-calibrated SPIRE observations. We highly recommend going through the recipe provided in \S 3 of \cite{Lombardi14} for the details of the calibration process. Repeating the procedure is beyond the scope of this paper. However, in brief, the procedure consists of the following steps:

\begin{enumerate}
    \item For each cloud, extract optical depth, temperature, and spectral-index map from the $Planck$ Legacy Archive\footnote{https://pla.esac.esa.int/\#home} for regions corresponding to $Herschel$ emission maps.
    \item For each $Herschel$ passband, calculate the expected emission assuming a modified blackbody model to make fiducial $Herschel$ maps from $Planck$ maps obtained in step 1.
    \item Degrade the original $Herschel$ emission maps to match the resolution of $Planck$-derived fiducial $Herschel$ map.
    \item Linear fit the relation between the original $Herchel$ maps with fiducial $Herschel$ maps for each wavelength. The offset of the linear fit provides the extended flux filtered out of the original $Herschel$ maps.
\end{enumerate}

\subsubsection{Modified blackbody fits} \label{obs:greybody_sg}

In this section, we briefly explain the procedures we employ to derive the column density and temperature maps and suggest that readers refer to \S2 of \cite{Pokhrel16} for more details. First, we matched all the $Herschel$ observations to a common resolution and grid that are equivalent to the 500 $\micron$ SPIRE map. Thermal dust emission is modeled by a blackbody spectrum that is modified by a frequency-dependent emissivity \citep{Hildebrand83}. Assuming that the dust emission is optically thin in the far-IR region, emission $I_{\nu}$ for a modified blackbody spectrum can be approximated as:

\begin{equation} \label{eq:greybody}
    I_{\nu} = \kappa_{\nu_0}(\nu/\nu_0)^{\beta}B_{\nu}(T)\Sigma ,
\end{equation}

\noindent
where $\kappa_{\nu_0}$ is the dust opacity per unit gas and dust mass at a reference frequency $\nu_0$. We took $\kappa_{\nu_0}$ = 2.90 cm$^2$/gm for $\nu_0$ corresponding to the longest observed wavelength, 500 $\micron$, following the OH-4 model \citep{Ossenkopf94}. $\beta$ is the dust emissivity power-law index, $B_{\nu}(T)$ is the Planck function for a perfect blackbody of temperature $T$, and $\Sigma$ is the mass surface density which is defined as $\Sigma$ = $\mu m_{H}N(H_2)$ where $\mu$ is the mean molecular weight per unit hydrogen mass $\sim$ 2.8, $m_H$ is the mass of single hydrogen atom and $N(H_2)$ is the gas column density. We assumed the canonical gas-to-dust ratio of 100 \citep{Predehl95} for converting dust measurements to gas.

\cite{Pokhrel16} utilized SPIRE maps to make column density and temperature maps, which we follow for the clouds in this study too. For Cygnus-X, however, the combined SPIRE 250 $\micron$ map suffers from a poorly matched background in different individual regions of the mosaic, and we could not use the SPIRE 250 $\mu$m map for our analysis of this cloud. Instead, we used the PACS 160 $\micron$ map, along with SPIRE 350 and 500 $\micron$ for studying dust/gas properties in the Cygnus-X star-forming complex. Since the majority of the 160 $\micron$ emission is from the warmer region and we include longer SPIRE bands as well for covering the colder regions, the final results of this study are not affected by substituting the 160 $\micron$ data for the 250 $\micron$ data.

\begin{figure}
\centering
\includegraphics[scale=0.39]{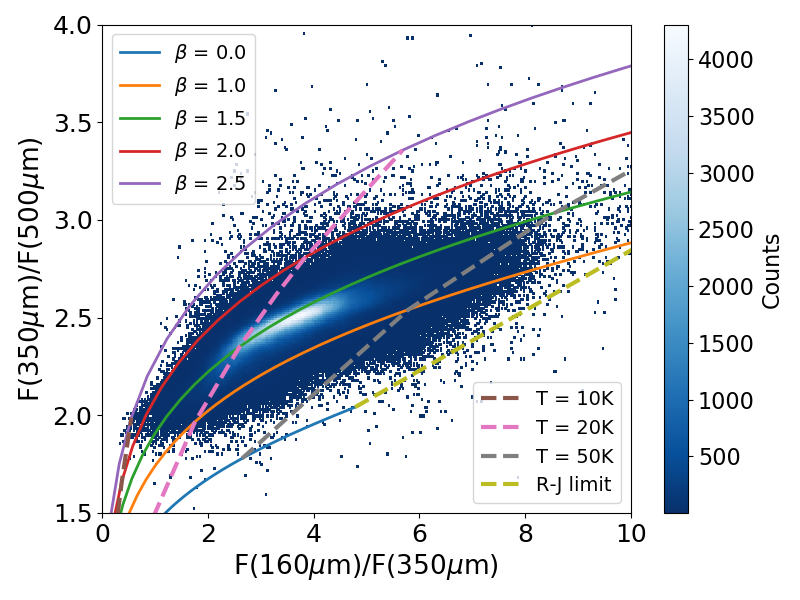}
\includegraphics[scale=0.39]{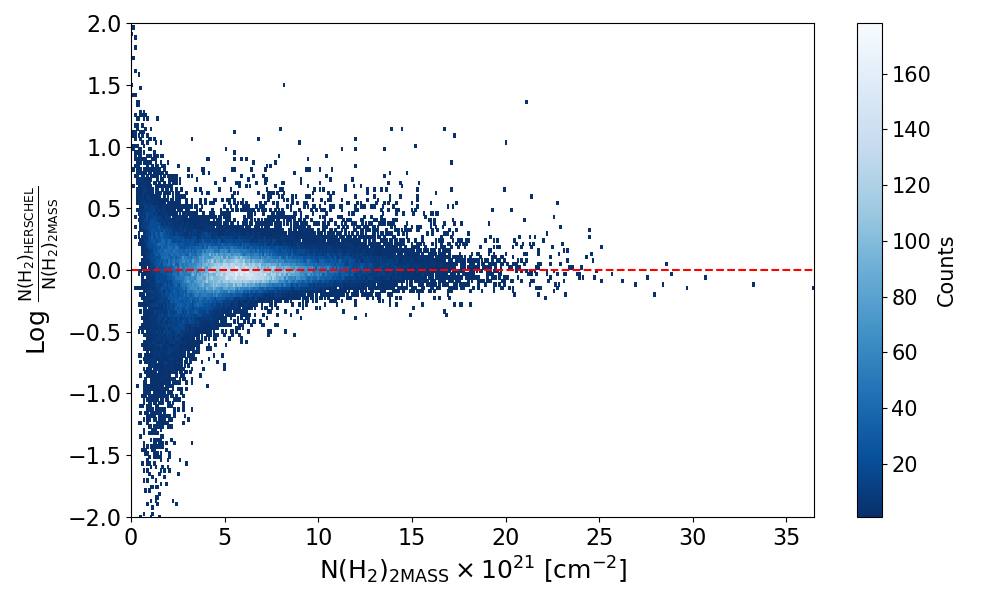}
\caption{Left: Flux ratio plot for the Cygnus-X star forming complex, showing the variation of $F_{350}/F_{500}$ with $F_{160}/F_{350}$ as a 2D histogram. Theoretical greybody  model showing different $\beta$ and temperature tracks are overplotted. The distribution in the flux ratio plot is best represented by $\beta$=1.5. Right: Comparison between the column density values obtained from dust emission with $Herschel$ (this work) and from extinction maps using 2MASS and UKIDSS \citep{Gutermuth11}. The distribution of the log of the ratio of the two column densities peaks at 0, showing a consistency in the column density values obtained by these two different methods.
}
\label{fig1}
\end{figure}

Equation \ref{eq:greybody} contains three unknown parameters: dust emissivity $\beta$, dust temperature $T$, and H$_2$ column density N(H$_2$). We used flux ratio plots similar to \cite{Pokhrel16} to find a representative $\beta$ that is typical in the cloud. The left panel of Figure \ref{fig1} represents such a plot for the Cygnus-X cloud complex that shows that $\beta$ = 1.5 is a representative value for Cygnus-X. Similarly, for other clouds, we constrained $\beta$ between 1.5 and 1.8. After fixing $\beta$, we performed a modified black body fits in three $Herschel$ wavebands using equation \ref{eq:greybody}, to obtain the column density and temperature maps.

Temperature estimation using $Herschel$ emission depends on the enclosure of the peak emission by their Spectral Energy Distribution (SED). Pixels with high-temperature estimates are prone to lie on the Rayleigh-Jeans (R-J) tail of the modified black body spectrum, rendering those estimates unconstrained toward higher values. This also affects column density estimation, as underestimating (overestimating) temperature in modified black body fits overestimates (underestimates) column density (see \citealt{Pokhrel16}). We examine the color-color space to find the pixels where emission may be R-J limited and exclude such pixels from further analysis. Table \ref{tab:cloudinfo_sg_fit} shows the list of emissivities and band ratios that we utilize for greybody fits, along with the R-J limited temperature for the cloud.

\begin{deluxetable}{lccccccc} 
\tablecaption{Greybody fit and YSO content for clouds \label{tab:cloudinfo_sg_fit}}
\tablehead{\colhead{Clouds} & \colhead{Band} & \colhead{$\beta$} & \colhead{T$_{\rm{RJ}}$} & \colhead{N$_{\rm{YSO, total}}$} & \colhead{N$_{\rm{Class I}}$} & \colhead{N$_{\rm{Class II}}$} & \colhead{$\frac{\rm{N_{Class II}}}{\rm{N_{Class I}}}$}\\ 
\colhead{ } & \colhead{($\mu$m)} & \colhead{ } & \colhead{$\mathrm{(K)}$} & \colhead{ } & \colhead{ } & \colhead{ } & \colhead{ }\\
\colhead{(1)} & \colhead{(2)} & \colhead{(3)} & \colhead{(4)} & \colhead{(5)} & \colhead{(6)} & \colhead{(7)} & \colhead{(8)}}
\startdata
Ophiuchus & (160, 250, 350, 500) & 2.0 & 27 & 351 & 70 & 281 & 4.01 \\
Perseus & (160, 250, 350, 500) & 2.0 & 27 & 452 & 100 & 352 & 3.52 \\
Orion-A & (160, 250, 350, 500) & 2.0 & 40 & 2394 & 294 & 2100 & 7.14 \\
Orion-B & (160, 250, 350, 500) & 2.0 & 40 & 544 & 91 & 453 & 4.98 \\
Aquila North & (160, 250, 350, 500) & 2.0 & 40 & 403 & 67 & 336 & 5.01 \\
Aquila South & (160, 250, 350, 500) & 2.0 & 35 & 911 & 160 & 751 & 4.69 \\
NGC 2264 & (250, 350, 500) & 1.7 & 26 & 558 & 100 & 458 & 4.58 \\
S140 & (250, 350, 500) & 1.5 & 35 & 531 & 61 & 470 & 7.71 \\
AFGL 490 & (250, 350, 500) & 1.5 & 35 & 319 & 45 & 274 & 6.09 \\
Cep OB3 & (250, 350, 500) & 1.8 & 24 & 2188 & 205 & 1983 & 9.67 \\
Mon R2 & (250, 350, 500) & 1.8 & 26 & 931 & 165 & 766 & 4.64 \\
Cygnus-X & (160, 350, 500) & 1.5 & 40 & 21387 & 2152 & 19235 & 8.94
\enddata
\tablecomments{(1) Molecular clouds that we investigate in this study (2) $Herschel$ wavebands used for making column density and temperature maps. (3) The value of the emissivity index chosen for Greybody fits. (4) Temperature above which the far-IR emission is consistent with Rayleigh-Jeans emission. (5) Total number of YSOs in the cloud. (6) The number of protostars (or Class I objects) in the cloud. (7) The number of YSOs (or Class II objects) in the cloud. (8) The ratio to Class II to Class I objects.}
\end{deluxetable}

The final step is a sanity check for the $Herschel$ derived column density maps. The column densities obtained using extinction maps are temperature independent, providing a valuable check of our fits to the dust emission. We compare our $Herschel$ derived gas column density values to near-IR extinction maps from \cite{Gutermuth11} and \cite{Rapson14} and find a reasonable agreement in all the clouds. Some of the extinction maps like Cygnus-X are unpublished but are made with the same technique as those in \citet{Gutermuth11}. The right panel of Figure \ref{fig1} shows the comparison between the column density map obtained using our method and the extinction based method for the Cygnus-X star-forming complex and shows a reasonable agreement. Plots for other clouds look similar so we do not show them.

\subsection{SESNA YSO catalog}

Astronomers have been using $Spitzer$ observations to identify and classify the young stellar systems for more than a  decade (e.g., \citealt{Gutermuth04, Harvey07, Allen07}). The method consists of comparing the mid-IR excess emission to the expected photospheric spectral energy distributions (SEDs) and colors of reddened photospheres and young stellar objects and using the slopes of the SEDs or colors to distinguish different types of YSOs. $Spitzer$ has surveyed over 90\% of known molecular clouds in the nearest 1 kpc and provided some fundamental results about the YSOs. See \cite{Evans09}, \cite{Kryukova12} and \cite{Dunham15} for results regarding the protostellar evolution, \cite{Gutermuth09}, \cite{Bressert10} and \citet{Megeath16} for demographics of YSO clustering, and \citet{Heiderman10}, \cite{Lada10} and \citet{Gutermuth11} for the $Spitzer$ results on gas-origins of YSO clustering.

 YSO catalogs made with different techniques and by different groups differ to various degrees. The initial examinations of YSO tallies derived for the same targets in c2d/GB Legacy Surveys \citep{Evans09, Dunham15} and \cite{Gutermuth09} suggested a 10-20\% discrepancy in the total count of YSOs was typical. \citet{Megeath16} found that the YSO incompleteness increases with the density of the YSOs, primarily due to the presence of bright nebulosity in dense, embedded clusters. Full catalog cross-matching reveals larger disagreements (Gutermuth et al. in prep.). YSO recovery rates also vary by technique and evolutionary class, leading to substantial systematic differences in the ratio of the number of pre-main-sequence stars with disks to protostars, a useful star-forming region evolutionary indicator.

Hence, while $Spitzer$-based surveys of star-forming clouds were a revolutionary step forward in their simultaneous extremely wide coverage and excellent mass completeness to dusty YSOs, many of these surveys were analyzed by independent groups that emphasized differing primary science goals and demonstrate clear discrepancies when compared. SESNA is a uniform retreatment of 92 (+16 for extragalactic contamination) sq. deg. of archival $Spitzer$ cryomission surveys of nearby star-forming regions to mitigate the discrepancies seen in previous $Spitzer$ YSO surveys. SESNA combines $Spitzer$ observations with the Two Micron All Sky Survey (2MASS) observations to cover a range of 1-24 $\mu$m to identify and classify YSOs with dusty circumstellar material. For the Cygnus-X star-forming complex, SESNA also uses deeper UKIDSS (UKIRT Infrared Deep Sky Survey; \citealt{Lawrence07}) near-IR Galactic Place Survey \citep{Lucas08}.

The SESNA YSO catalogs are relatively uniform in their observing parameters and data treatment including mosaic construction, source extraction algorithms, photometric measurement techniques, and source classification technique. In Table \ref{tab:cloudinfo_sg_fit} we include the number of Class I and Class II objects in the SESNA catalog for different clouds. More details of the catalog with the public release of the YSO census will be included in an upcoming paper (Gutermuth et al. in prep). For the purpose of this paper, we denote the protostars as Class I and all more evolved YSOs with disks as Class II objects.

\subsubsection{Edge-on disk contamination} \label{edgeon_sg}

When using mid-IR photometry in classifying the Class II YSOs, there is a chance of misclassifying edge-on Class II as Class I \citep{Crapsi08, Offner12}. Radiative transfer modeling shows that most flat spectrum YSOs can be explained as inclined pre-main-sequence stars with disks as opposed to protostars (e.g., \citealt{Robitaille07}). \cite{Crapsi08} argued that of an ensemble of pre-main-sequence stars with disks, $>$39\% would be confused as flat-spectrum sources if classified by the slope between their fluxes at 2 and 24 $\micron$. \citet{Furlan16} used 4.5-2.4 $\micron$ photometry including the IRS spectra between 5-30 $\micron$ and found that out of 321 protostars in Orion including 102 $Spitzer$ identified flat-spectrum sources, only 4 lacked evidence for envelopes. This showed that the $Spitzer$ identified protostar sample is not strongly contaminated by more evolved YSOs.

The SESNA catalog utilizes 1 - 24 $\micron$ photometry in classifying YSOs and protostars. The technique emphasizes a color criteria that are demonstrably less susceptible to reddening or edge-on classification confusion \citep{Gutermuth09}. Still, any YSO class ambiguity can yield a large systematic uncertainty in the Class II and Class I source counts and their ratios. We require an accurate Class II to Class I ratio for this work so we correct for the Class I ambiguity caused by edge-on Class II.

\cite{Gutermuth09} estimated edge-on disk confusion in two well populated and relatively evolved young clusters, assuming that all protostars identified in those regions must be edge-on disk contaminants. They found that the likelihood of confusion from inclined disks is small, 3.6\% $\pm$ 2.6\% (2/56) for IC 348 Core-1 and 3.1\% $\pm$ 1.8\% (3/96) for IC 5146 Core respectively. In agreement with this, assuming all the protostars in the off-cloud region of the Cep OB3b cluster with edge-on disks, \cite{Kryukova12,Kryukova14} found that 2-5\% of protostars were misclassified edge-on disks for many clouds including Cygnus-X. Thus, if all those protostars are indeed edge-on Class IIs, then $<$1/30 Class II may be misidentified as Class I due to an edge-on disk orientation. This corresponds to a Class II/ Class I ratio of 30, whereas the median value for the ratio in their survey is 3.7. Hence, to remove the possible effect of edge-on contamination on the ratio of Class II and Class I, we use the estimate that edge-on disks contaminate the Class I protostar tally by 3.5\% of the Class II tally. The assignment is performed statistically in terms of the total count of each class of YSOs, rather than individually.

\subsubsection{Extragalactic contamination}

Another source of contamination is residual background extragalactic contamination. The most common contaminants are Active Galactic Nuclei (AGN) that mimic the SEDs of flat spectrum YSOs (see \citealt{Gutermuth09, Grossschedl19}). Most broad-line AGN are removed with a color-magnitude cut, however, some residual AGNs still contaminate the YSO population. Based on the SESNA analysis of 16 sq. deg. of $Spitzer$ archival observations of two so-called ``blank" fields used for extragalactic studies, Bo$\rm{\ddot{o}}$tes and Elias-N1, there are 9 $\pm$ 1 residual extragalactic contaminants per sq. deg. in the SESNA YSO catalogs. These are nearly uniformly divided between the Class I and Class II SED classes. For a given area of a cloud, we estimate the number of possible AGN candidates in that area and remove that from the total number of each class of YSOs.

\subsubsection{YSO completeness correction for Cygnus-X} \label{cygx_corr_sg}

For our study, we select large clouds with projected area $>$100 pc$^2$ and containing $>$100 YSOs with their local YSO surface densities ranging by at least an order of magnitude. Although a uniform sample of YSOs, the varying distance to the clouds cause slightly different levels of mass completeness. Also, the presence of bright sources and bright, structured nebulosity near the centers of some clusters can alter the completeness relative to regions of lower YSO density \citep{Gutermuth11,Megeath16,Gutermuth15}. The incompleteness effect can also be observed for localized regions of clouds with significantly high surface densities such as ONC and NGC 2024 in the Orion molecular cloud \citep{Megeath16}, but we mask high-density pixels with A$_{\rm{V}}$ $>$ 100 mag so they will have a minimal effect in our analysis.

The YSO sensitivity for the SESNA catalog is $\sim$0.1 M$_{\odot}$ for the clouds that are $<$1 kpc distance. The sensitivity drops for more distant clouds. For Cygnus-X, field stars are denser, regions of bright nebulosity are more common, and the IRAC data are shallower (3/4 of the integration time of the closer cloud surveys). Thus, the YSO sensitivity in Cygnus-X is intrinsically lower ($\sim$1 M$_{\odot}$). For Cygnus-X, we correct the number density of YSOs to make a uniform sensitivity catalog assuming an initial mass function (IMF) characterization.

We adopt the group IMF of \citet{Chabrier03} to estimate a correction for the missing low mass YSOs in Cygnus-X. We calculate the integrated IMF for two values of minimum mass, 0.1 M$_{\odot}$ and 1 M$_{\odot}$, and a maximum of 150 M$_{\odot}$. The fraction, $\frac{\int_{1M_{\odot}}^{150M_{\odot}}IMF}{\int_{0.1M_{\odot}}^{150M_{\odot}}IMF}$ gives the fraction of YSOs that are detected in the SESNA catalog for the Cygnus-X star-forming complex. We find this fraction to be 0.163. Thus, we divide $\Sigma_{*}$ by this fraction to get the corrected population down to the sensitivity of 0.1 M$_{\odot}$, equivalent to the other clouds in our sample.

\section{Results}
\label{sec:nnd_sg}

\subsection{Calculation of $\Sigma_{*}$ and $\Sigma_{\rm{gas}}$}

We implement an $n^{th}$ nearest neighbor surface density analysis of the SESNA YSO catalog in each cloud. The analysis is similar to the one employed in e.g., \cite{Gutermuth09}, \cite{Gutermuth11}, \cite{Bressert10}, \cite{Megeath16} and \cite{Li19} where the nearest neighbor distance ($d_n$) is calculated as the distance between a particular YSO and its $n^{th}$ nearest neighbor. A YSO in a less crowded region will cover a bigger area (mean separation to the $n^{\rm{th}}$ neighbor is larger) than a YSO in a more crowded region, but the number of YSOs in both areas are equal. The mass surface density for YSO is calculated as (c.f. \citealt{Casertano85}):

\begin{equation} \label{eq:sigstar}
    \Sigma_{*} = \Bigg ( \frac{n - 1}{\pi d_n^2} \Bigg ) M_{\rm{*}}.
\end{equation}

\noindent
The mean mass of YSOs, $M_{*}$, is assumed to be 0.5 M$_{\odot}$ \citep{Evans09}. To assess the effect of YSO clustering at different smoothing scales, we perform the nearest neighbor analysis for $n$ = 4, 6, 11 \& 18, similar to \cite{Gutermuth08} and \citet{Sokol19}. The fractional uncertainty in $\Sigma_{*}$ is ($n$ - 2)$^{-0.5}$. Higher values of $n$ result in poorer spatial resolution but smaller fractional uncertainty \citep{Casertano85, Gutermuth11}. 

We use k-dimensional trees (\citealt{Maneewongvatana99}, or k-d trees in short) to find the nearest neighbor distances for four different values of $n$. A k-d tree is a binary tree where each node specifies an axis and splits the set of data based on whether their coordinate along that axis is greater than or less than a particular value. The axis and splitting points are chosen by the ``sliding midpoint" rule, which ensures that the cells do not all become long and thin. The tree can be queried for the $n$ closest neighbors of any given point. We first convert the YSO positions to Cartesian coordinates, then use k-d trees to find $d_n$ and hence $\Sigma_{*}$ using Equation \ref{eq:sigstar}.

We calculate the corresponding gas mass surface density $\Sigma_{\rm{gas}}$ using $Herschel$ derived gas column density maps. To directly compare the mass surface densities of YSO and gas, we sample the gas density at areas enclosed by a circle of radius $d_n$ centered on each YSO position. The average column density $\overline{N(H_2)}$ is then converted to $\Sigma_{\rm{gas}}$ using the following relation (c.f. \citealt{Bohlin78, Gutermuth11}):

\begin{equation} \label{conversion}
    \Sigma_{\rm{gas}} = \overline{N(H_2)} \frac{15}{0.94 \times 10^{21}} ~\rm{M_{\odot}/pc^2}.
\end{equation}

\begin{figure}
\centering
\includegraphics[scale=0.35]{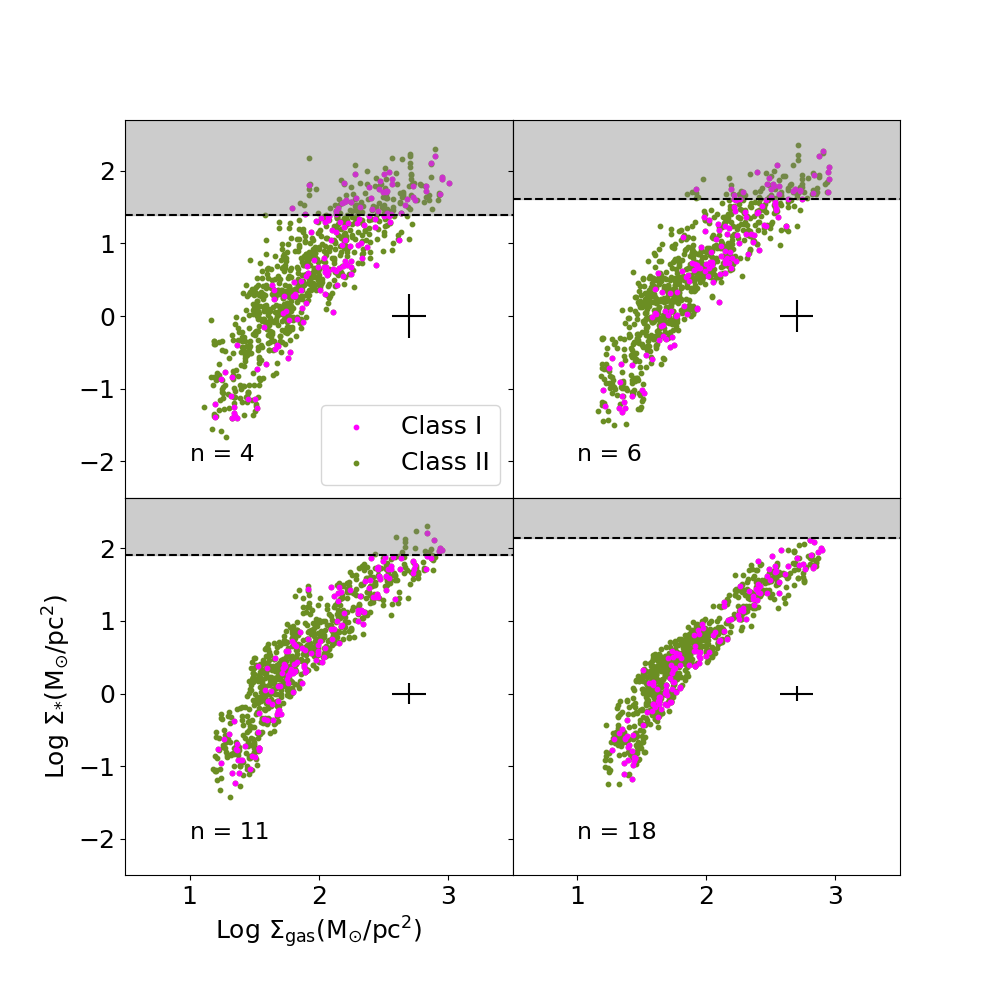}
\includegraphics[scale=0.35]{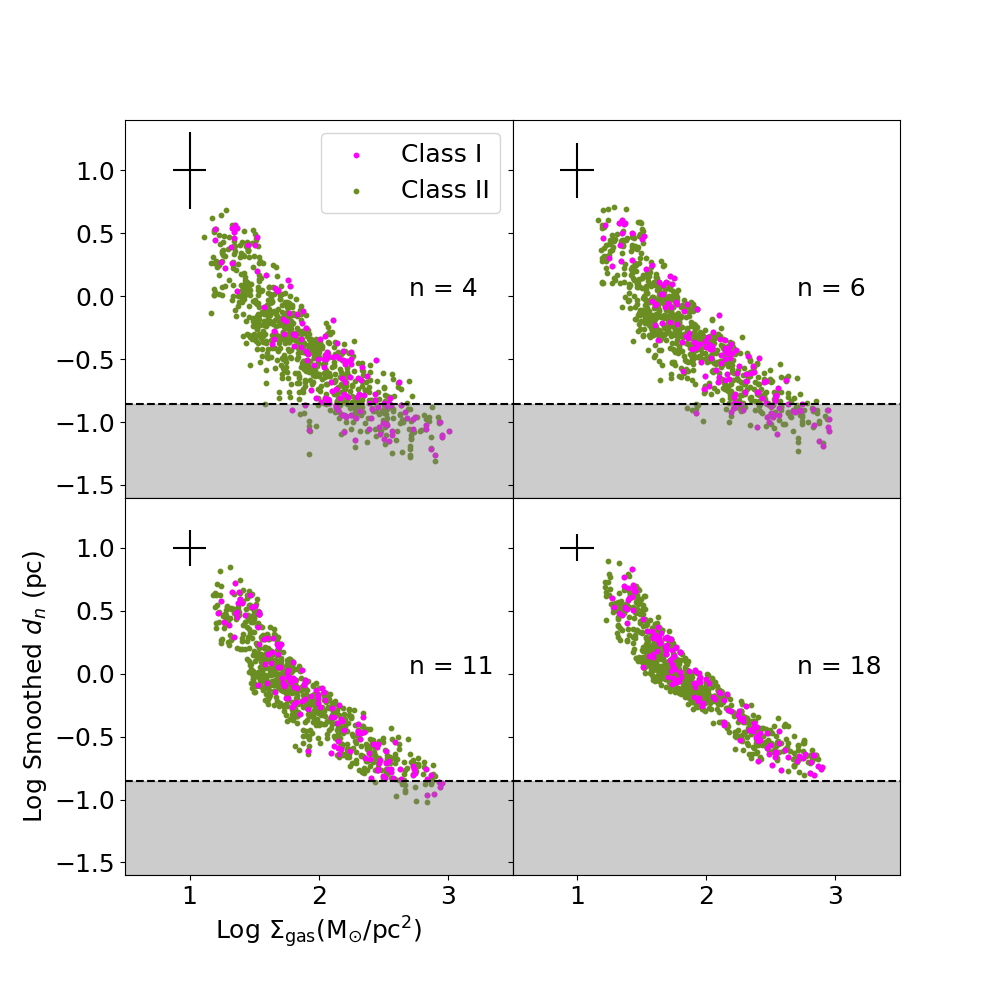}
\caption{Plots showing the systematic effect of varying $n$ in our measured quantities for the Mon R2 cloud. Left panel shows the change in $\Sigma_{*}$ with $\Sigma_{\rm{gas}}$ for $n$=4, 6, 11 and 18. Right panel shows the variation of distance to the $n^{\rm{th}}$ neighbor with $\Sigma_{\rm{gas}}$ of region enclosed by the circle with radius $d_{\rm{n}}$. The black cross represents typical uncertainties. Grey shaded areas show the representation of the $Herschel$ resolution limit of 36$''$ in the smoothing size that is set by the stellar clustering in Mon R2.
}
\label{stargas_monr2}
\end{figure}

The left panel of Figure \ref{stargas_monr2} shows the star-gas surface density correlation using different choices of $n$ for the Mon R2 GMC. Similarly, the right panel of Figure \ref{stargas_monr2} shows the variation of smoothed distance to the $n^{th}$ neighbor ($d_{n}$) with $\Sigma_{\rm{gas}}$ in the region enclosed by a circle of radius $d_{n}$. Uncertainties in $\Sigma_{*}$ are calculated based on their fractional uncertainty, whereas the typical uncertainty in $\Sigma_{\rm{gas}}$ is $\sim$30\% (based on the column density uncertainty analysis for Mon R2 in \citealt{Pokhrel16}). The range of $d_{n}$ for Mon R2 GMC varies from $\sim$0.05 pc to $\sim$5 pc for $n$ = 4, and from $\sim$0.15 pc to $\sim$8 pc for $n$ = 18. The grey shaded area in the right panel of Figure \ref{stargas_monr2} shows the region limited by the SPIRE 500 $\micron$ resolution of 36$''$. Similarly, the grey region in the left panel of Figure \ref{stargas_monr2} corresponds to the SPIRE 500 $\micron$ resolution limit in the smoothing size that is set by the stellar clustering using Equation \ref{eq:sigstar}. 

Figure \ref{stargas_monr2} shows that the higher column density regions have lower $d_{n}$, thus they are more densely populated with YSOs. As we increase $n$, the smoothing size scale increases to include larger areas and thus lower spatial resolution. Each panel in Figure \ref{stargas_monr2} shows a similar locus of points and a clear correlation between our star and gas density measures, as reported in \citet{Gutermuth11}. We do not find any change in the character of the locus for 4 $\leq$ $n$ $\leq$ 18. For $n$=11, the regions within the $Herschel$ resolution limit consist $<$2\% of the overall stellar densities for Mon R2. All other clouds, except Cygnus-X, are closer than Mon R2. Hence, their resolution limit corresponds to even fewer stellar densities in the grey region. For Cygnus-X, after the IMF correction, we found $<$2\% stellar densities in the grey region, similar to Mon R2. Hence, for all further analysis, we select $n$=11 because of its good compromise between the smoothing size, uncertainty and $Herschel$-equivalent resolution limit.

\subsection{Star-gas surface density correlations}

\begin{figure}
\centering
\includegraphics[scale=0.43]{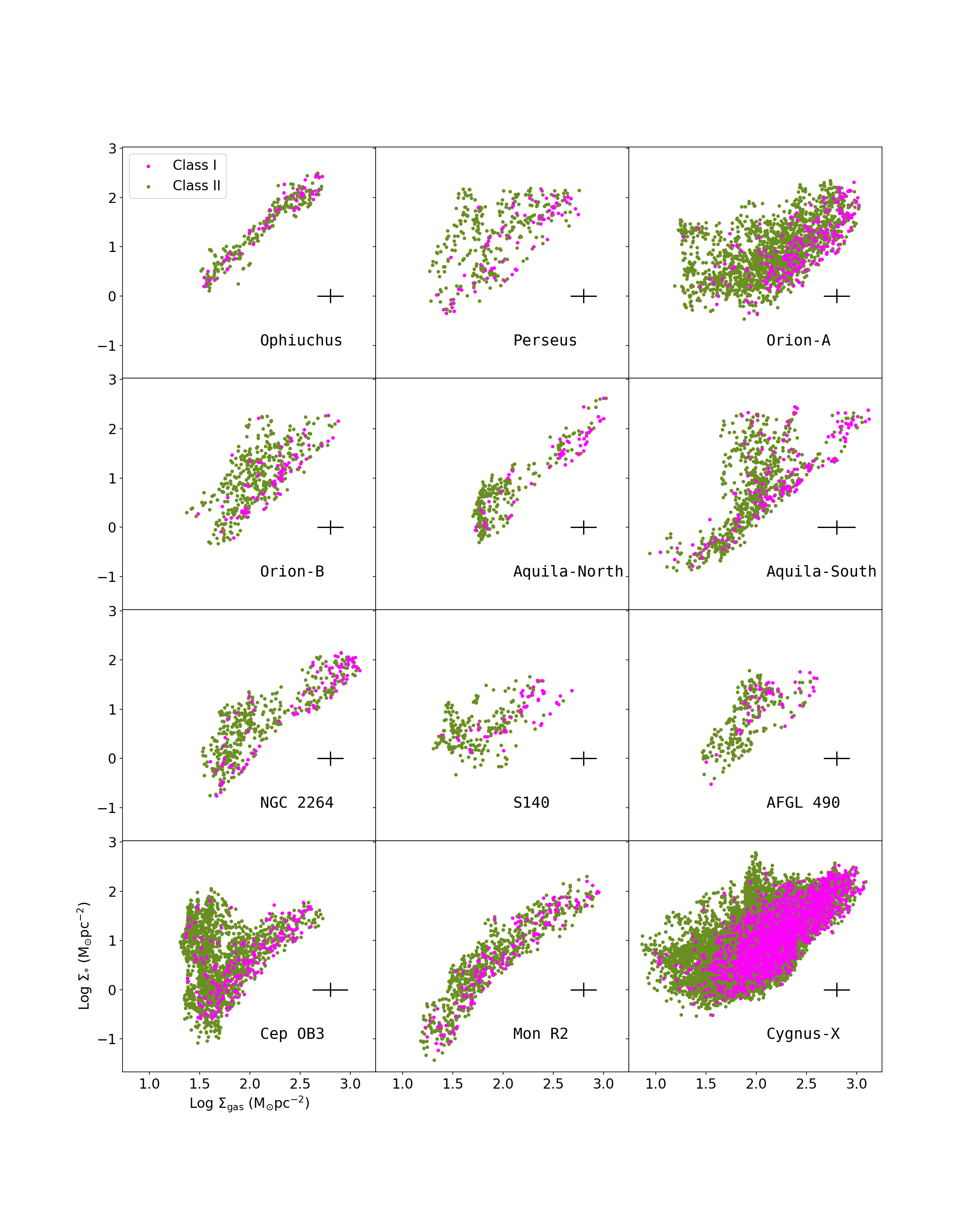}
\vspace{-0.8in}
\caption{YSO mass surface density versus gas mass surface density for our sample of clouds. The YSO mass surface density is calculated for the nearest 11$^{\rm{th}}$ neighboring YSOs sampled at each YSO, and the gas surface density is derived from the corresponding identical area in the column density map. Class I objects are shown in magenta and Class II are shown in light green. The black cross in each panel shows the typical error bar. 
}
\label{stargas_first}
\end{figure}

There are ample studies in the past that show a spatial alignment of YSOs in projected dense gas structures \citep{Megeath04, Gutermuth05, Gutermuth08, Gutermuth09, Gutermuth11, Allen07, Evans09, Lada13, Zari16,Lada17, Li19}. These studies show that most of the clouds contain a higher concentration of YSOs in the regions with higher gas densities. To quantify the apparent correlation between the distribution of the YSOs and gas, in Figure \ref{stargas_first} we plot the stellar mass surface density $\Sigma_{*}$ versus the gas mass surface density $\Sigma_{\rm{gas}}$ for twelve molecular clouds. In each of these, we measure the surface and gas density at the position of a known YSO. The markers are colored to distinguish densities centered on protostars and more evolved stars with disks.  

Figure \ref{stargas_first} shows a star-position-sampled star-gas density relation for both Class I and Class II. The observed star-gas surface density correlations in Figure \ref{stargas_first} can be empirically divided into three types. The first type (Type-A) is defined by a single, distinct star-gas surface density correlation locus (only a primary branch). Examples of this Type are Ophiuchus, Aquila North, NGC 2264 and Mon R2. The second type (Type-B) is similar but includes further correlation branches (secondary branches) in addition to the primary branch. We assign Perseus, Orion-B, Aquila South, S140, AFGL 490 and Cep OB3 to Type-B. The third kind (Type-C) does not have a clear primary branch and exhibits a much wider span of points in the plot space. The two largest star-forming clouds in our sample, Orion-A, and Cygnus-X fit this third type. This variety of morphological types has been reported before \citep{Gutermuth11}, but our uniform datasets and analysis give us sufficient confidence in these differences to analyze them in more detail.

\subsubsection{Regional evolution based on CII/CI ratio}

\citet{Gutermuth11} reported that the morphological variations in star-gas surface density correlations are due to YSO build-up over time, gas dispersal by outflows and non-coevality in each cloud. The presence of non-coeval regions in the clouds is confirmed by variations in the ratio of the numbers of Class II to Class I (designated as CII/CI throughout the paper). Based on the histograms of $\Sigma_*$/$\Sigma_{\rm{gas}}^2$ for both Classes,  \citet{Gutermuth11} reported a common evolutionary trend between 3 $\times$ 10$^{-4}$ and 5 $\times$ 10$^{-3}$ pc$^2$ M$_{\odot}^{-1}$ that contained greater than half of all YSOs with excess IR emission in every cloud. The approach was useful for explaining the different morphology in the star-gas surface density correlations. We take a similar approach in this study by defining regional age variances based on the CII/CI ratio.

\begin{figure}
\centering
\includegraphics[scale=0.4]{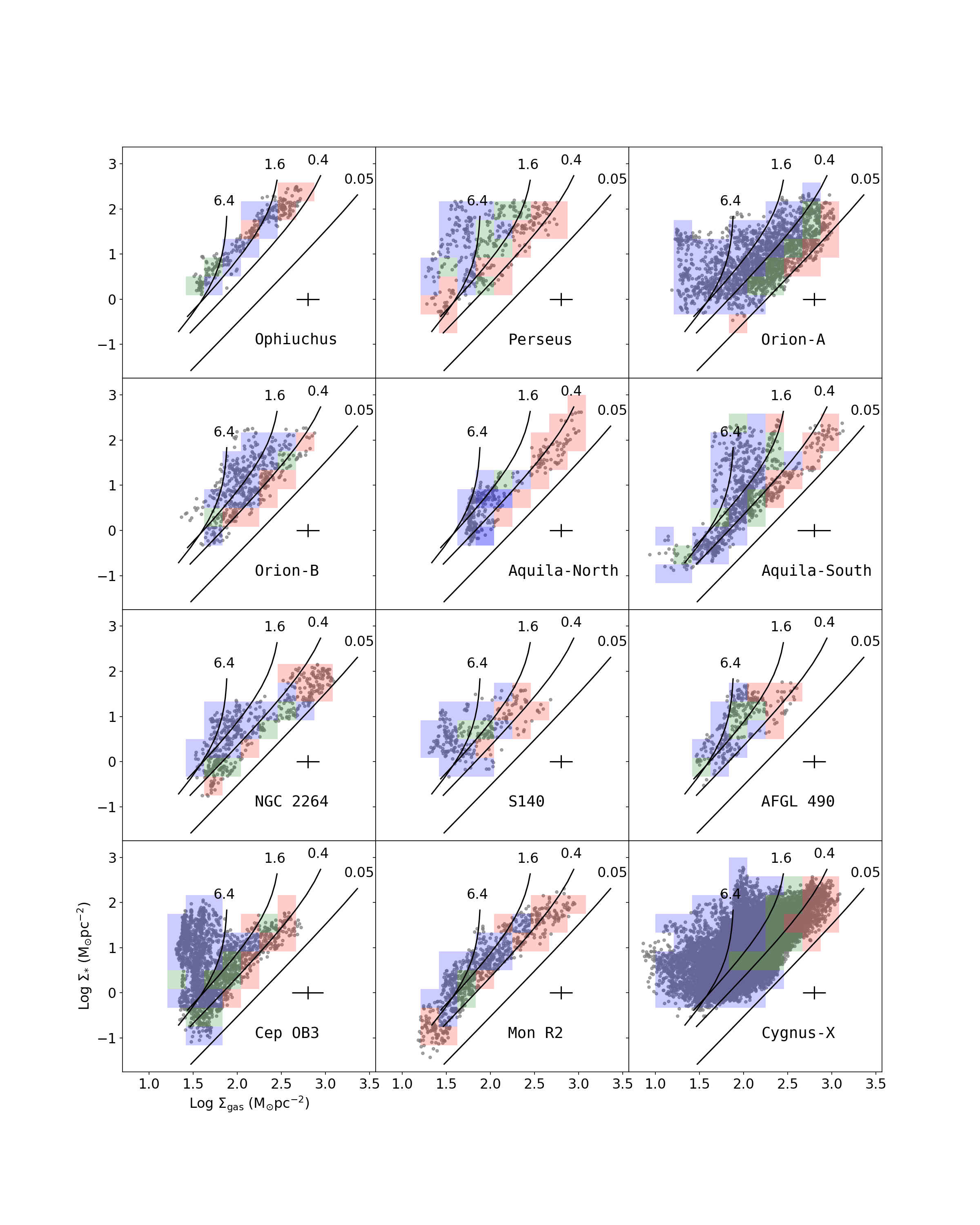}
\vspace{-0.5in}
\caption{Surface density plots shown in Figure \ref{stargas_first}, now overlaid with model isochrones and color-coded to indicate relative YSO evolution. Uniform grids are created in the surface density data and are color-coded based on the ratio of Class II to Class I. Red cells have CII/CI $<$ 3, green cells have 3 $<$ CII/CI $<$ 10, and blue cells have CII/CI $>$ 10. Black isochrone lines represent the star-gas locus at a particular evolutionary time in Myr (discussed in \S\ref{model_sg}).
}
\label{stargas_model_sg}
\end{figure}

Figure \ref{stargas_model_sg} is similar to Figure \ref{stargas_first} but adds color-coded cells representing the typical evolutionary age in that region of the star-gas density space, based on CII/CI ratio within each cell. Black solid lines represent isochrones of a model that we will discuss in detail in \S\ref{model_sg}. We will focus on the colored cells only in this section.
Each panel is uniformly sampled into grid cells in such a way that the spacing between the grids in $\Sigma_{\rm{gas}}$ is $\sim$0.2 M$_{\odot}$/pc$^2$ and in $\Sigma_*$ it is $\sim$0.4 M$_{\odot}$/pc$^2$. The grid sampling is uniform across all the clouds.

The color in Figure \ref{stargas_model_sg} represents the prevailing YSO evolutionary stage for the given region of the star-gas density space. \citet{Gutermuth11} provides the range of CII/CI values for the stellar population at different evolutionary stages. In Figure \ref{stargas_model_sg}, we used such CII/CI values to color-code the population at different evolutionary stages. The grid cells with CII/CI $<$ 3 represent the youngest population in the group and are colored red. If 3 $<$ CII/CI $<$ 10, the cells are colored green indicating an intermediate evolutionary stage with a mixture of young (protostars) and old (disks) stellar sources. If CII/CI $>$ 10, the box is colored blue and they represent the oldest evolutionary stage in the group. Thus, in Figure \ref{stargas_model_sg}, red cells are protostar-rich with a very recent rise in their star formation rates, green cells contain more evolved populations than red cells, and blue cells contain the most evolved YSO population.

Most of the red cells in Figure \ref{stargas_model_sg} lie in the higher $\Sigma_{*}$ and $\Sigma_{\rm{gas}}$ regimes, indicating high star formation rate and a predominantly young stellar population in the densest parts of molecular clouds. Similarly, at the lower end of the star-gas locus with lower $\Sigma_{*}$ and $\Sigma_{\rm{gas}}$ contains a mixture of different colored cells, indicating multiple star formation epochs. Regions that have higher $\Sigma_{*}$ and lower $\Sigma_{\rm{gas}}$ are predominantly blue, indicating the most evolved regions with low current star formation and a higher concentration of older YSOs. In \S \ref{finalplots_sg}, we constrain the correlation indices based on the regional evolutionary state and after correcting for contamination.

\subsection{Constraining the underlying power-law index} \label{finalplots_sg}

Given the brief protostellar lifetime, the initial condition of star formation is imprinted in Class I, while Class IIs may be evolved and impacted by the external environment in some clouds. Hence, Class I are used widely in the literature to study star-gas surface density correlations \citep{Heiderman10, Evans14, Willis15} to reduce uncertainty in the ages of the stars and their proximity to their birth sites at the expense of substantially reduced source counts. For our final version of the star-gas surface density correlation plots, we take an intermediate path by centering our measurements on Class I but measuring the combining densities of Class I and Class II. Using Figure \ref{stargas_model_sg}, we find highly evolved regions with CII / CI $>$ 30 (see \S\ref{edgeon_sg}) and exclude them. In such regions, the population of protostars may be dominated by edge-on disk contamination.  By excluding these regions, we reduce the influence of edge-on disk contamination. This ensures that we only sample regions with ongoing star formation when determining the underlying star-gas surface density correlation.

\begin{figure}
\centering
\vspace{-0.5in}
\includegraphics[scale=0.43]{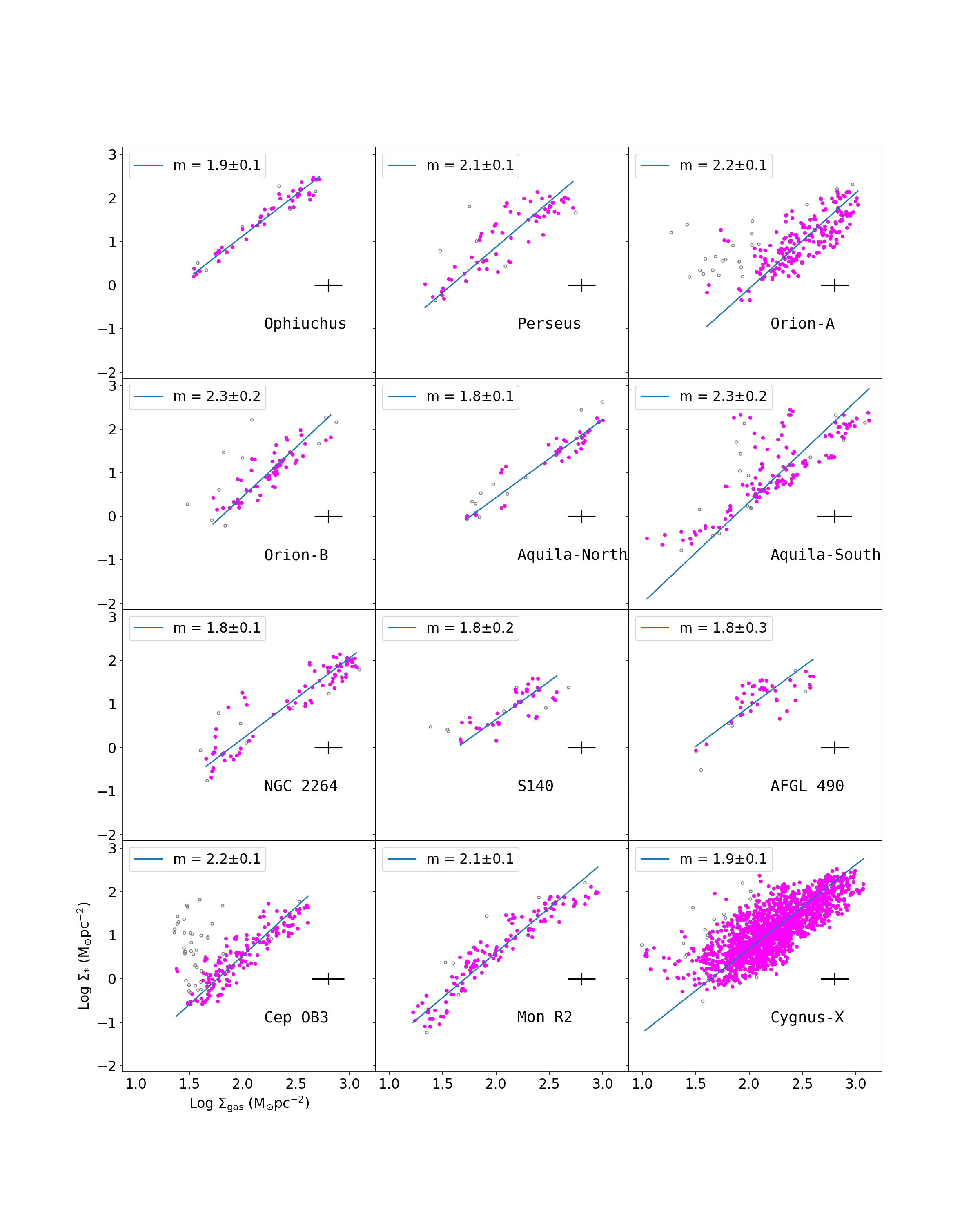}
\vspace{-0.5in}
\caption{Surface density plots for the YSOs and gas for Class I protostars. Following Figure \ref{stargas_model_sg}, the data that belong to the cells where CII/CI $>$ 30 represent highly evolved YSO clusters that do not trace active, star-forming regions. Such data are shown as grey open circles. 
The remaining data where CII/CI $<$ 30 are shown as magenta circles. The best-fit values for magenta data are noted in each panel, the average of which is $\sim$2.
}
\label{stargas_fits}
\end{figure}

Figure \ref{stargas_fits} shows the resulting star-gas surface density correlation, showing the variation of $\Sigma_*$ with $\Sigma_{\rm{gas}}$ for Class I objects. The regions with CII/CI $>$ 30 are older and more evolved regions, usually that have undergone gas dispersal. 
The protostars in the regions with CII/CI $<$ 30 are shown as magenta circles and represent regions with active, ongoing star formation. The linear fits in each panel are obtained using the Orthogonal Distance Regression method, by taking account of uncertainty along both axes. The linear best-fit equations are presented in column 2 of Table \ref{tab:rchi2}. The best-fit power-law indices range between 1.8 and 2.3, with an average index of $\sim$2. The range of power-law indices in \citet{Gutermuth11} is between 1.4 and 3.8. Thus, with $Herschel$ observations and SESNA YSO catalog, we obtain a much narrower distribution of power-law indices. The consistency of the power-law index in our analysis of star-gas surface density correlations is a remarkable result, considering that our sample of clouds has a varying range of mass, size, age, average star formation rate, and peak star formation rates.

Note that $\Sigma_{*}$ in Figure \ref{stargas_fits} can be readily converted into $\Sigma_{\rm{SFR}}$ by dividing by the average lifetime of the YSOs. The correlation is the same when plotting in terms of $\Sigma_{\rm{SFR}}$ and $\Sigma_{\rm{gas}}$. However, to be consistent with literature that implements the nearest neighbor technique (e.g., \citealt{Gutermuth11}), we plot Figure \ref{stargas_fits} in terms of $\Sigma_*$.

The inferred power-law index shows that stars form more efficiently at high column densities than at low column densities. As the cloud evolves, gas at high $\Sigma_{\rm{gas}}$ depletes faster and the star-gas surface density correlation steepens. The steeper star-gas power-law index is consistent with the stellar build-up and local gas mass depletion at higher column density locations. Below, we present a semi-analytic model of star formation based on the depletion of available gas.

\subsection{Gas depletion model with disk decay} \label{model_sg}

With the identification of a consistent power-law star-gas locus in all twelve clouds, we confirm one assumed aspect of the interpretive model of \citet{Gutermuth11}, hereafter referred to as the G11 model, namely that $\Sigma_{\rm{SFR}} \propto \Sigma_{\rm{gas}}^2$.  Now we revisit that model and refine it to improve agreement with our observational constraints.

\citet{Gutermuth11} presents a simple semi-analytic model of star formation and gas depletion to explain the different branches observed in Type-B and Type-C star-gas surface density correlation plots that can be seen in Figure \ref{stargas_model_sg}. The G11 model predicts that different branches in the star-gas density space are caused by regional cloud evolution. However, that model needs adjustment to describe our new data to take into account disk evolution.

The foundation of the G11 model is a star formation law where the star formation rate per area varies with the power-law of the local instantaneous gas mass surface densities. They made the following assumptions to simplify and analytically solve the model.

\begin{enumerate}[i)]
    \item The YSOs do not move significantly from their birth site to within the scale of our measurements ($\sim$1 pc scale).
    \item The molecular gas is not flowing into or within the molecular cloud at $\sim$1 pc scale; gas is either being converted into a stellar mass or is being ejected via outflows.
    \item There is no effect on the parsec-scale gas distribution by large-scale feedback such as from supernova or local winds and radiation from high-mass stars. 
\end{enumerate}

The first assumption is supported by two observations. Even in dense configurations, many YSO groupings show relatively little dynamical evolution \citep{Gutermuth09}, and the typical initial velocity dispersion of YSOs is small ($\lesssim$0.4 km/s; \citealt{Walsh07, Muench07}). It is also supported by simulations of clusters forming in turbulent clouds \citep{Offner09,Kirk14}. The second assumption is adopted for simplicity and may not be true for smaller scales $<$1 pc with gas infall (e.g., \citealt{Walsh06,Kirk13}). The third assumption is based on the fact that such large scale processes affect the gas distribution at longer timescales ($>$10 Myr) than we are considering, with the exception of high-mass star feedback, which we ignore for simplicity. These processes may also be more influential on GMC formation (e.g., atomic to molecular transition). 

The model adopts a general star formation law where the star formation rate density is a power-law function of the gas mass surface density:

\begin{equation} \label{eq:model_sg_sfl}
    \frac{\partial \Sigma_*}{\partial t} = ck\Sigma_{\rm{gas}}^\alpha
\end{equation}

\noindent
where $\Sigma_*$ and $\Sigma_{\rm{gas}}$ are the mass surface density of YSOs and gas respectively, $\alpha$ is the power-law dependence, $c$ is the mass conversion efficiency that accounts for how much mass is ejected ($M_{\rm{ejected}}$) from outflows and winds and defined as $c = \frac{M_*}{M_* + M_{\rm{ejected}}}$, and $k$ is the gas depletion rate which is defined as $\frac{\partial \Sigma_{\rm{gas}}}{\partial t} = -k \Sigma_{\rm{gas}}^\alpha$. An extensive analytical solution for the model is provided in G11. In Figure \ref{stargas_fits}, we show that the underlying power-law index for star-gas surface density correlation for the entire diversity of clouds in our sample is $\sim$2. So, we proceed with $\alpha$ = 2 in the G11  model. Equation \ref{eq:model_sg_sfl} can then be reduced to,

\begin{equation} \label{eq:model_sg}
    \frac{\Sigma_*}{\Sigma_{\rm{gas}}^2} = ckt\Bigg( 1 + \frac{t}{t_0} \Bigg)
\end{equation}
where $t_0$ is the gas depletion time scale (see \citealt{Gutermuth11} for details). 

The stellar density obtained by infrared observations refers to the YSO population that has circumstellar disks. Optically thick disks are known to disperse with time, as the fraction of stars with disks declines with increasing mean stellar age (e.g., \citealt{Haisch01,Hernandez08}). Typically the fraction of young stars with optically thick primordial disks and/or those which show spectroscopic evidence for accretion appear to approximately follow an exponential decay with characteristic time $\sim$2.5 Myr or a half-life of $\approx$ 1.7 Myr (see \citealt{Mamajek09} for details, especially their Figure 1). 

The G11  model does not consider the exponential decay of disks. Looking back to Figure \ref{stargas_model_sg}, $Herschel$ observations show a mixed population at the lower $\Sigma_{*}$ and lower $\Sigma_{\rm{gas}}$ region, which can be explained in terms of disk decay of YSOs at higher evolutionary age. In this work, we add the recipe for disk decay in the G11  model for estimating stellar density. This change is modeled as $\Sigma_*^{\rm{obs}}$ = $\Sigma_* \rm{exp}(\frac{-t}{t_{\rm{disk}}})$, where t$_{\rm{disk}}$ is the characteristic time scale for disk decay as discussed above, and $\Sigma_*$ is the total number of stars formed in time $t$. With this addition to the model, the YSO populations with $>$2.5 Myr age exhibit decreasing stellar density at lower $\Sigma_{*}$ and $\Sigma_{\rm{gas}}$ values. This effect causes an overlap with less evolved isochrones. This agrees with the observed mix of regions at lower gas and stellar density with intermediate and varying CII/CI ratio.

\subsection{Comparing model with observed star-gas surface density plots}
\label{comparision}

The model isochrones for different evolutionary age are overplotted in Figure \ref{stargas_model_sg}. Different realizations of Equation \ref{eq:model_sg} are considered for star formation ages of 0.05, 0.4, 1.6 and 6.4 Myr. We have adopted a mass conversion efficiency $c$=0.3, gas depletion rate $k$=0.002 Myr$^{-1}$pc$^2$M$_{\odot}^{-1}$, and t$_{\rm{disk}}$=2.5 Myr. These values are similar to \cite{Gutermuth11} but are tuned to match the ages inferred from the ratio CII /CI in our cloud sample, the inclusion of disk decay, and also for better high column density sensitivity and dynamic range of the $Herschel$ observations. The assumed values are also consistent with recent simulations studying dense core efficiencies \citep{Offner14,Offner17}. In \citet{Gutermuth11}, two out of eight clouds show a power-law index of $\sim$2, each with a Type-A star-gas morphology. In this analysis, we find a power-law index of $\sim2$ for all twelve clouds in our sample. Thus, Type-A star-gas morphology is described by a power-law index of 2, and all clouds have evidence of a Type-A-like locus centered on their Class I populations.

The G11 model predicts the steepening in some of the secondary branches seen in Type-B star-gas morphology, but it fails to explain the overlap at low to medium $\Sigma_{\rm{gas}}$. Such an overlapping feature is prominent in Type-B morphology clouds, such as Aquila-South. The robustness of the $Herschel$ data ensures higher confidence in such features in Type-B clouds, thereby justifying the addition of disk decay to the G11  model. In Figure \ref{stargas_model_sg}, after adding the disk decay in the G11 model, we see the isochrones overlap at low to medium $\Sigma_{\rm{gas}}$ as in observed star-gas surface density correlations.
 
Type-C star-gas morphology is like Type-B but so extreme in terms of the number of star-forming events that they have many secondary branches that overlap and blur the distinct loci. Both Type-C clouds in our sample, Orion-A and Cygnus-X have high-mass stars and evidence of strong regional feedback as well. The model is not built to handle such feedback, so this is left for future work. Feedback may shift $\Sigma_{\rm{gas}}$ to the left in Figure \ref{stargas_model_sg}, as gas disperses, leaving stars behind and reducing the gas column density relative to the star density. 

The star-gas surface density correlation isochrones from the model for different evolutionary ages show that different branches seen in the star-gas surface density correlation plots can be explained by the systematic evolution of the star-gas system from the initial correlation. Given that several assumptions go into the model and there are uncertainties associated with the observed star-gas surface density correlation locus, the model does a good job of explaining the observed regional evolution in the correlation plots. More work is required to incorporate feedback effects that are still very poorly understood (e.g., \citealt{Krumholz14}).

\section{Discussion} \label{sec:discussion_sg}

\subsection{Star-gas observations as constraints for models}

Stars form as an end product of the hierarchical, multi-scale fragmentation of molecular clouds (c.f. \citealt{Pokhrel18}). The observed star-gas density correlation for the recently formed stars provides strong new constraints on the models of cloud fragmentation and star formation. Below, we present a comparison of the observed to the synthetic star-gas density correlations for two models. The first model is a simple analytical model of the fragmentation of a molecular cloud that is supported only by thermal gas motions, with a typically observed geometry and under simple assumptions. The second model is a more complex turbulent hydrodynamic simulation of molecular gas with self-gravity that includes radiative feedback from the protostars.

\subsubsection{Predictions from an analytic model of thermal fragmentation} \label{jeansfrag_sg}

Recent observations of cloud morphology in the nearby molecular clouds support sheet-like geometry \citep{Beaumont10, Arce11,Qian15}. Modulated layers of gas are (meta)-stable and fragment like a uniform sheet, thereby allowing for the wider range of column densities that are observed in nearby clouds \citep{Myers09}. We model the formation of stars due to fragmentation of the dense gas in an isothermal, self-gravitating layer of gas cloud (sheet-like geometry).  From \cite{Larson85}, the Jeans mass in such a system is given by,
\begin{equation} \label{eq:MJ}
    M_J = \frac{A \sigma^4}{\Sigma_{\rm{gas}}G^2}
\end{equation}
where $A$ = 4.67 for an isothermal, self-gravitating layer of gas \citep{Larson85}, $\sigma$ is the gas velocity dispersion, $\Sigma_{\rm{gas}}$ is the gas mass surface density, and $G$ is the gravitational constant.

The number surface density of Jeans masses is defined as,
\begin{equation} \label{eq:NJ}
    N_J = \frac{\Sigma_{\rm{gas}}}{M_J}
\end{equation}

We assume that one Jeans mass of cloud forms one star, i.e., $N_J$ = $N_*$, where $N_*$ = $\Sigma_*$/$M_*$ and $M_*$ = 0.5 M$_{\odot}$. Combining Equations \ref{eq:MJ} and \ref{eq:NJ}, we get the dependence of $\Sigma_*$ over $\Sigma_{\rm{gas}}$ as,

\begin{equation} \label{eq:thermaljeanspred}
    \Sigma_* = \frac{1}{A}\Bigg(\frac{0.5 \rm{M_{\odot}}~G^2}{\sigma^4}\Bigg) ~ \Sigma_{\rm{gas}}^2
\end{equation}

\noindent
For the case of thermal fragmentation, $\sigma$ in Equation \ref{eq:thermaljeanspred} is the same as the sound speed. Using an average temperature of the cloud to estimate $\sigma$, a simple Jeans fragmentation in a self-gravitating, isothermal layer of gas predicts that if the initial structure-inducing physics of YSO clustering is consistent with the thermal fragmentation of a modulated sheet gas geometry \citep{Myers09}, the star-gas surface density correlations will have a power-law index of 2.

We note that Equation \ref{eq:thermaljeanspred} represents an over-simplified fragmentation scenario, where the gravitational contraction is opposed solely by thermal gas motions that are dictated by a single temperature, and assuming that stars form in a single event of fragmentation. Furthermore, it assumes that the mass of the formed star is independent of the Jeans mass and the surrounding gas density. These assumptions are certainly not true in the molecular clouds. Hence, more detailed modeling of clouds is required to capture the fragmentation mechanism in the clouds. Such an attempt is described below, where we model fragmentation with hydrodynamic simulations.

\subsubsection{Predictions from a turbulent hydrodynamic simulation} \label{hdturbulence_sg}

We use simulations from \citet{Qian15} to explore the star-gas density correlation in the clouds that are dominated by hydrodynamic turbulence. For a fair comparison with $Herschel$-derived star-gas density correlation results in Figure \ref{stargas_fits}, we use the simulations to obtain $Herschel$-like synthetic observations. Below we briefly explain the simulation and synthetic observations and refer to \citet{Offner09} for details of the numerical methods.

The simulations are performed using the {\tt ORION} adaptive mesh refinement (AMR) code \citep{Truelove98,Klein99}. We use a simulation with periodic boundary conditions, with 256$^3$ base grids and 4 AMR levels of refinement. The domain size for simulations is 5 pc, contains $\sim$3780 M$_{\odot}$ gas with an initial temperature of 10 K. The simulation solves for the gas and radiation temperature by solving the flux-limited diffusion radiative transfer equation. The turbulence is initialized by adding random velocity perturbations with an input power spectrum, $P(k)$ $\propto$ $k^{0}$ for $k$ = 1 $\sim$ 2. A turbulent steady state is obtained when the power spectrum and density distribution function are constant in time. To obtain a steady state, the perturbations are injected for two domain crossing times without self-gravity. Turbulence decays due to dissipation in shocks over a crossing time (e.g., \citealt{Stone98,MacLow99}). Hence, we continue injecting energy so that the velocity dispersion remains constant even after turning on gravity in the system. Once the clouds begin collapsing, refinement is added in the simulation box so that the Jeans criterion is satisfied for a Jeans number of $N_{\rm{J}}$ = 0.125 \citep{Truelove97}. When the density exceeds the Jeans resolution on the maximum AMR level, a star forms and we insert a sink particle in that position \citep{Krumholz04}. The stars adopt a sub-grid model for radiative feedback, which takes into account accretion luminosity and protostellar evolution \citep{Offner19}. We take the simulation snapshot at $\sim$1.27 Myr after the gas self-gravity is turned on. By this time, most of the protostars have already formed, but they have not migrated over parsec scales from their birth sites.

To produce synthetic observations, we flatten the snapshot gas density cubes onto a fixed grid to obtain a 2D density map. The projected density maps are given an arbitrary world coordinate system based on Field04 in Mon R2 in \citealt{Sokol19} and are scaled to the resolution necessary for the synthetic observation to appear at a given distance. The process is followed by the reprojection of projected 2D density maps into $Herschel$-SPIRE 500 $\mu$m maps. Finally, we convert the projected density in the synthetic maps to the molecular column density maps by dividing by the product of the mass and mean molecular weight of hydrogen. Similarly, the sink particle positions are also flattened and projected onto the same world coordinate grid. Finally, since the simulations have periodic boundary conditions, we tile the projected density map (both for the gas and sink particles) to fill Field04 in \citet{Sokol19} to mitigate the effects of the edge-of-field-based bias in the individual simulation snapshots.

The resolution of the sink particles in simulations is finer than the resolution of young stellar objects in $Spitzer$ observations. We match the resolution of the sink particles with observations by blending the sink particles that are closer than the resolution of the observed YSOs using $Spitzer$. First, we estimate the nearest separation distance for all the sink particles. If the separation of a pair of sink particles is less than minimum observed distance, we average their position with an average mass of 0.5 M$_{\odot}$. In this way, all the pairs of sink particles are averaged to blend them in the first iteration. The process is followed by the second iteration where again the sink particles that are closer than observed minimum separation are blended. The process is repeated until the minimum separation of a pair of sink particles matches the minimum separation of observed young stellar sources. The process is repeated for the observed clouds and synthetic observation at all the different distances of clouds and corresponding simulations.

\subsubsection{Comparison between observations and model/simulation predictions}

\begin{figure}
    \vspace{-0.8in}
    \centering
    \includegraphics[scale=0.47]{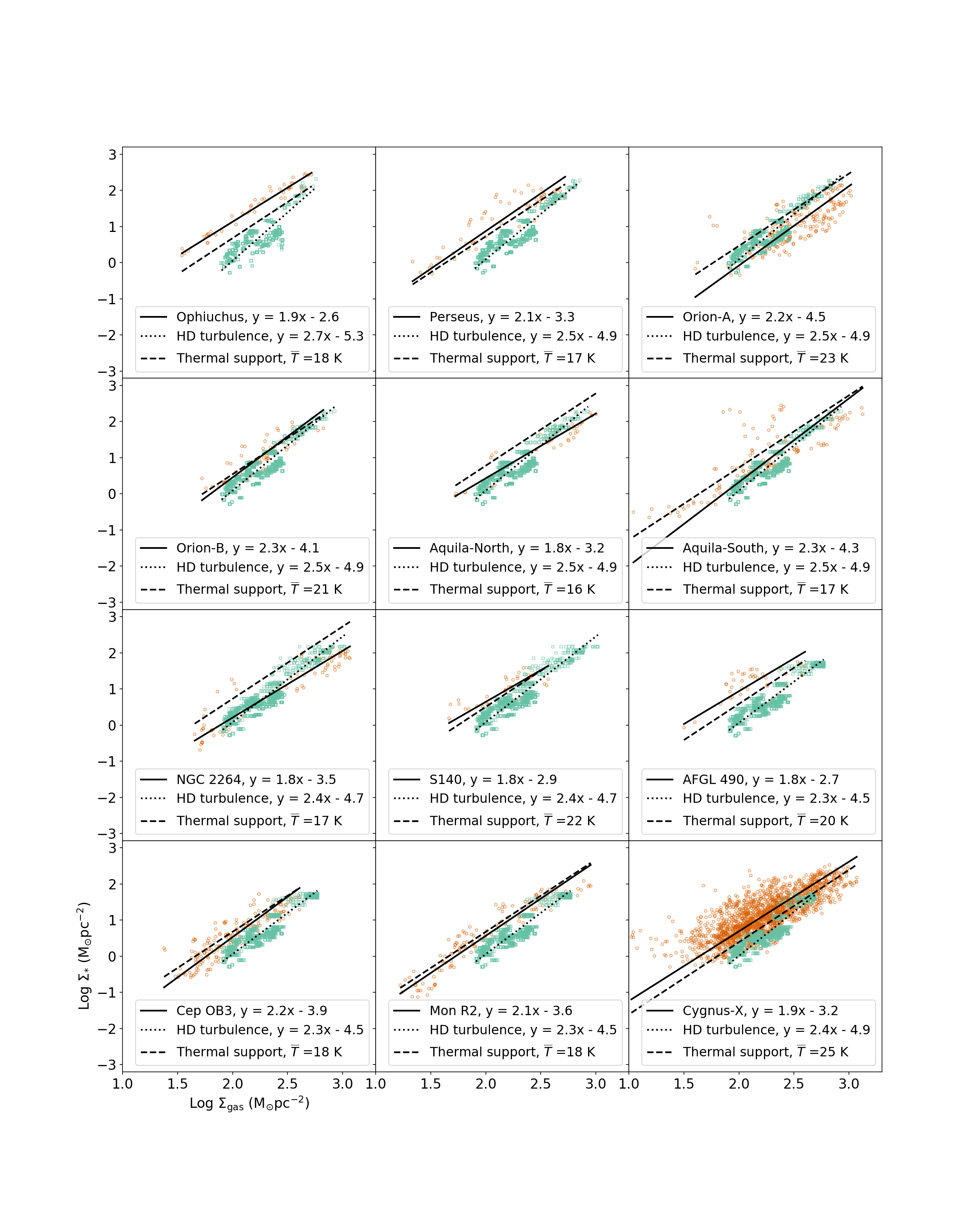}
    \vspace{-1.0in}
    \caption{Comparison of the observed star-gas density correlations for different clouds with two theoretical predictions. First, using hydrodynamic turbulent simulations, and second, using Equation \ref{eq:thermaljeanspred} for thermal Jeans fragmentation scenario. The temperatures used for predicting the thermal support is the average temperature at the YSO positions. The prediction from the thermal Jeans fragmentation is shown as a dashed line. The  best-fit linear equation (in log plots) for observation (brown circles) and simulation (green squares) are shown as the solid and dotted black line, respectively.}
    \label{fig:stargas_sim}
\end{figure}

Figure \ref{fig:stargas_sim} shows a comparison between the observed star-gas density correlation and the two theoretical predictions that are described in \S \ref{jeansfrag_sg} and \ref{hdturbulence_sg}. The actual observations are plotted as open brown circles, with their best-fit line in log-axes plotted as the solid black line. For the case of the thermal Jeans fragmentation, we use a single-valued temperature in Equation \ref{eq:thermaljeanspred} to compute $\Sigma_*$ for a range of values of $\Sigma_{\rm{gas}}$. The range of $\Sigma_{\rm{gas}}$ is chosen to be the same as for observations. The temperature map is obtained by the modified black body fits of $Herschel$ maps (see \S \ref{subsec:herschel_sg}). We used the average temperature at the YSO positions in the cloud to estimate $\Sigma_*$ in Equation \ref{eq:thermaljeanspred}. The prediction of the Jeans thermal fragmentation is overplotted as the black dashed line. Similarly, the simulated data is represented by the green open squares, with their best-fit line shown by the black dotted line. The average temperatures used for thermal support and the best-fit linear equations are mentioned in the legend.

We implement the chi-squared technique as a goodness-of-fit parameter to compare observed and model/simulated correlations. Table \ref{tab:rchi2} presents reduced chi-squared values between the thermal support model (Column 4 in Table \ref{tab:rchi2}) and hydrodynamic simulations (Column 5 in Table \ref{tab:rchi2}) with observations. The reduced chi-squared values show a consistency of the thermal support model for Ophiuchus, Perseus, Orion-B, Aquila-South, S140, AFGL 490, Cep OB3, Mon R2, and Cygnus-X. Similarly, predictions with hydrodynamic turbulent simulation are more consistent with Orion-A, Aquila-North and NGC 2264. We emphasize that the hydrodynamic simulation used in this study represents only one realization of a possible set of synthetic results, and hence the reduced chi-squared analysis is not used for a decisive preference of one model over the other.

Column 6 in Table \ref{tab:rchi2} lists the average temperature at the YSO positions that is used to infer thermal support in Equation \ref{eq:thermaljeanspred}, the prediction for which is presented in Figure \ref{fig:stargas_sim}. Column 7 in Table \ref{tab:rchi2} lists the temperature required for the thermal support to be consistent with observations. The average temperature in Ophiuchus, Perseus, S140, AFGL 490 and Cygnus-X is more than the expected temperature for pure thermal support. This would imply that the temperature in these clouds has increased since the clouds fragmented due to radiation feedback. Similarly, the average temperature in Orion-A, Aquila-North, Aquila-South and NGC 2264 have lesser temperatures than that required for pure thermal support. Finally, Orion-B, Cep OB3, and Mon R2 have a consistency between their average temperature and that required for thermal Jeans fragmentation. Thus, in these clouds, the initial fragmentation temperature has remained intact after the stars formed. The dynamics of the stellar systems and gas kinematics have not changed the fragmentation conditions.

Equation \ref{eq:thermaljeanspred} assumes a single fragmentation event that forms stars in clouds, and it uses a single temperature to estimate $\Sigma_*$. These are oversimplified assumptions. Cloud temperature varies in different regions of the cloud and at different times and scales. The consistency is observed at a cost of no modeling of how the cloud happens to be structured as it is, nor any capacity to evolve the cloud further over time (to deal with the impact of mechanical feedback, for instance). Yet, the consistency of the star-gas density prediction from such a simplified model with several observed clouds is noteworthy.

\begin{deluxetable}{lcccccc}
\tablecaption{Reduced $\chi^2$ values for different models and temperature estimates.
\label{tab:rchi2}}
\tablehead{\colhead{Clouds} & \colhead{Best fit} & \colhead{Best fit} & \colhead{$\chi^2_{\rm{red,th}}$} & \colhead{$\chi^2_{\rm{red,sim}}$} & \colhead{$\overline{T}$} & \colhead{T$_{
\rm{th=obs}}$}\\
\colhead{ } & \colhead{ (Observation) } & \colhead{ (Simulation) } & \colhead{ } & \colhead{ } & \colhead{$\mathrm{(K)}$} & \colhead{$\mathrm{(K)}$}\\
\colhead{(1)} & \colhead{(2)} & \colhead{(3)} & \colhead{(4)} & \colhead{(5)} & \colhead{(6)} & \colhead{(7)}}
\startdata
Ophiuchus & y = 1.9x - 2.6 & y = 2.7x - 5.3 & 10 & 42 & 18 ($\pm$1) & 11 ($\pm$1) \\
Perseus & y = 2.1x - 3.3 & y = 2.5x - 4.9 & 7 & 35 & 17 ($\pm$1) & 14 ($\pm$1) \\
Orion-A & y = 2.2x - 4.5 & y = 2.5x - 4.9 & 17 & 16 & 23 ($\pm$5) & 40 ($\pm$1) \\
Orion-B & y = 2.3x - 4.1 & y = 2.5x - 4.9 & 3 & 8 & 21 ($\pm$3) & 21 ($\pm$1) \\
Aquila-North & y = 1.8x - 3.2 & y = 2.5x - 4.9 & 14 & 7 & 16 ($\pm$1) & 26 ($\pm$1) \\
Aquila-South & y = 2.3x - 4.3 & y = 2.5x - 4.9 & 15 & 24 & 17 ($\pm$2) & 26 ($\pm$1) \\
NGC 2264 & y = 1.8x - 3.5 & y = 2.4x - 4.7 & 23 & 11 & 17 ($\pm$1) & 33 ($\pm$1) \\
S140 & y = 1.8x - 2.9 & y = 2.4x - 4.7 & 5 & 17 & 22 ($\pm$1) & 19 ($\pm$1) \\
AFGL 490 & y = 1.8x - 2.7 & y = 2.3x - 4.5 & 13 & 40 & 20 ($\pm$2) & 13 ($\pm$1) \\
Cep OB3 & y = 2.2x - 3.9 & y = 2.3x - 4.5 & 5 & 14 & 18 ($\pm$2) & 21 ($\pm$1) \\
Mon R2 & y = 2.1x - 3.5 & y = 2.3x - 4.5 & 4 & 17 & 18 ($\pm$2) & 20 ($\pm$1) \\
Cygnus-X & y = 1.9x - 3.2 & y = 2.4x - 4.9 & 12 & 35 & 25 ($\pm$1) & 18 ($\pm$1)
\enddata
\tablecomments{(1) Molecular clouds that we investigate in this study. (2) Best-fit equations for observational data (c.f. Figure \ref{stargas_fits} \& \ref{fig:stargas_sim}), where y = log($\Sigma_{\rm{*}}$) and x = log($\Sigma_{\rm{gas}}$). (3) Best-fit equations for HD simulation data (c.f. Figure \ref{fig:stargas_sim}) where y = log($\Sigma_{\rm{*}}$) and x = log($\Sigma_{\rm{gas}}$).
(4) Reduced $\chi^2$ values for the thermal support model. (5) Reduced $\chi^2$ values for the turbulent hydrodynamic simulation. (6) The average temperature at the YSO positions, with uncertainties on the variation with the mass averaged temperature. (7) The temperature required for the cloud to fragment thermally, i.e., the temperature at which the model prediction for thermal support is consistent with the observed clouds. These temperature estimates are obtained with the least square minimization of temperature in Equation \ref{eq:thermaljeanspred}.
}
\end{deluxetable}

On the other hand, the turbulent hydrodynamic simulation incorporates multiple star formation events and captures more practical details of fragmentation than given by Equation \ref{eq:thermaljeanspred}. Our further analysis shows that correcting for the cloud geometry (following \citealt{Qian15}) provides further consistency in the correlation coefficient prediction from simulations for other clouds (see Appendix \ref{app:cropping}). To extend the star-gas density correlation predictions from simulations to other clouds requires information about the three-dimensional geometry of the gas clouds in our sample. Furthermore, a more direct comparison between the simulations and the observations requires the inclusion of magnetic fields and kinematic feedback. However, doing such simulations is beyond the scope of this paper.

\subsection{Fragmentation in Mon R2: cloud, cores and YSOs} \label{monr2_frag_sg}

Various studies report star formation as a consequence of multiscale hierarchical fragmentation of molecular clouds (e.g., \citealt{Wang14, Pokhrel18, Beuther19}). Analogous to the star-gas density studies, relations between core-gas densities can be used to study whether fragmentation of cores dictates the density of protostars. One such core-gas density correlation has been studied for Mon R2 by \cite{Sokol19} using the AzTEC/LMT identified cores and $Herschel$ gas map from \cite{Pokhrel16}.

\begin{figure}
    \centering
    \includegraphics[scale=0.54]{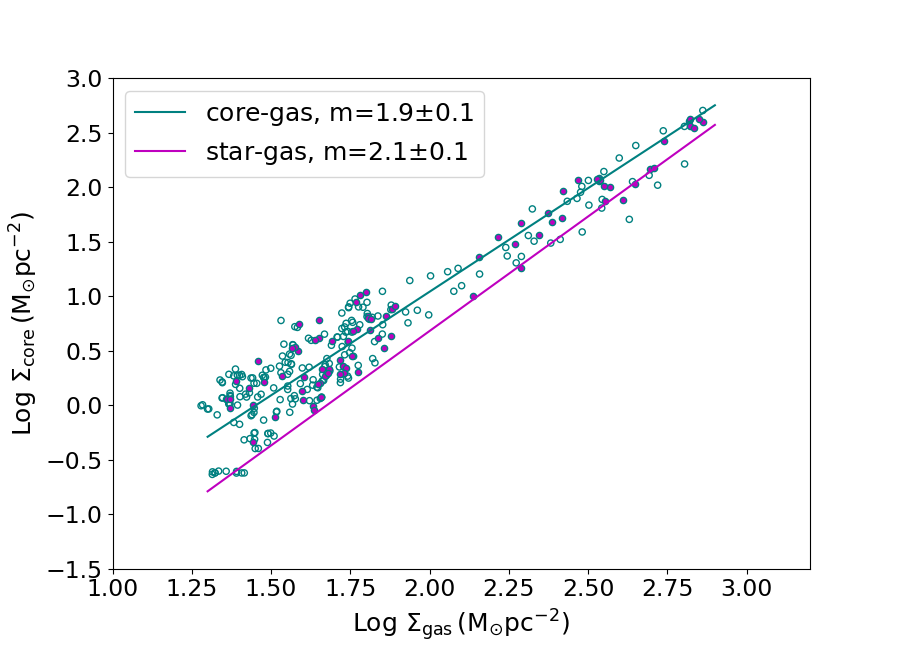}
    \caption{Core-gas correlations for Mon R2 from \citet{Sokol19}. Filled circles represent protostellar cores and open circles are starless cores. The best-fit linear line on the cores has a slope of 1.9 $\pm$ 0.1. Magenta line shows the best-fit slope of 2.1 $\pm$ 0.1 for star-gas surface density correlation for Mon R2 from our study. The offset between the best-fit lines for core-gas and star-gas correlations is $\sim$0.3 dex.}
    \label{fig:stargascore}
\end{figure}

\cite{Sokol19} used the nearest neighbor technique for $n$=11 to investigate core-gas density correlation in Mon R2 with core masses measured using AzTEC/LMT. We scaled their results for the core-gas correlations for $d$=860 pc and present them in Figure \ref{fig:stargascore}. The Figure shows that the best-fit linear slope between the logarithms of $\Sigma_{\rm{core}}$ and $\Sigma_{\rm{gas}}$ is 1.9 $\pm$ 0.1. In our study, we explored the star-gas surface density correlations for Mon R2 and found the best-fit linear slope of 2.1 $\pm$ 0.1 for $n$ = 11. This result is overplotted in Figure \ref{fig:stargascore} as the magenta line. The power-law index between core-gas correlation and star-gas surface density correlation for Mon R2 are consistent to within 1-$\sigma$. The similarity in the core-gas density correlation and star-gas density correlation for Mon R2 indicates that the initial conditions of fragmentation of clouds in forming cores have remained intact while forming stars. This suggests that the Mon R2 cloud is kinematically young enough that primordial structure has yet to be erased by dynamical interactions at $>$1 pc scales.

The offset between the best-fit locus lines for the core-gas and star-gas density correlations gives the mass efficiency for converting core gas mass into protostellar mass (e.g., $c$ parameter in the G11 model). For Mon R2, we found this offset to be 0.3 $\pm$ 0.1, which is consistent with our assumed value of $c$ in Equation \ref{eq:model_sg_sfl}, \S\ref{model_sg}. Also, the offset may represent the mass efficiency between the core mass function (CMF) and the stellar IMF. We note two caveats in this calculation. First, we assume a constant stellar mass to calibrate the stellar locus, which may not have a big impact as the mass variation in protostars are narrow enough to explain such a large offset. Second, \citet{Sokol19} assumed a constant temperature and dust emissivity to compute core masses, making the core masses uncertain up to a factor of two.

In Mon R2, the YSOs (both CI and CIIs) and cores (both protostellar and starless) are not just similarly clustered relative to their local diffuse gas density, they are also roughly co-spatial \citep{Gutermuth11, Pokhrel16, Sokol19}. Feedback from the protostars does not seem to have changed the spatial arrangement of the newly forming cores. Further core-gas density studies for other clouds are required to better understand the impact of protostellar feedback on core evolution in more detail. A follow-up study includes the Clouds-to-Cores (C2C) Legacy Survey with TolTEC/LMT that will investigate more than an order of magnitude core-gas density correlations across a varying environment. The corresponding star-gas density correlations for the clouds in the C2C survey are already explored in this study.

\subsection{Absence of the gas density threshold for star formation}

Among the twelve clouds we have studied here, we find no compelling evidence for a gas column density threshold for star formation in the range of $\Sigma_{\rm{gas}}$ $\sim$ 20 - 1000 M$_{\odot}$/pc$^2$, neither as a strict binary threshold nor as a change in the power-law index. This result is consistent with a growing tally of recent studies that refute the existence of a gas density threshold for star formation (e.g., \citealt{Gutermuth11, Krumholz12, Khullar19}). 

In the case of dense core formation, the threshold A$_{\rm{V}}$ for core detection varies from cloud to cloud and core statistics are limited by clustering and resolution. Hence, the generally lower number of cores in lower A$_{\rm{V}}$ as seen in \cite{Johnstone04,Konyves15} may be a direct effect of core-gas correlation with power-law index $\sim$2 \citep{Sokol19}. Likewise, lower numbers of YSOs in lower column density regions and a considerably larger number of YSOs in higher column density regions, such as clusters, is a direct manifestation of the super-linear star-gas surface density correlation that we explored in this study, rather than a particular threshold around which the star-gas characterization changes abruptly.

Our result that there is no threshold is based on the star-sampled nearest neighbor technique used to explore the star-gas density correlations. We have minimized the effect of edge-on disk contamination in our analysis (see \S \ref{edgeon_sg}) and considered only Class I protostars in our final results (c.f. Figure \ref{stargas_fits}) to minimize the impact of YSO migration. Some other literature studies that explore the existence of threshold are based on gas-sampled extinction contours (e.g., \citealt{Heiderman10,Lada12,Evans14}). To directly compare with their results, we need to implement their techniques in our observation. Examining such technique differences in detail is beyond the scope of this paper, but will be addressed in a future one.

One caveat regarding the absence of a column density threshold result is that the analysis presented in this study is confined to scales that are larger than dense cores. For example, for $\Sigma_{\rm{gas}}$ $\sim$ 50 M$_{\odot}$/pc$^2$, the structures with size scale of 0.4 - 2.9 pc with typical size of 1 pc are resolved in this study; for $\Sigma_{\rm{gas}}$ $\sim$ 100 M$_{\odot}$/pc$^2$, the structures of 0.25 - 1.2 pc size scale are resolved with typical resolved size of 0.6 pc; and for $\Sigma_{\rm{gas}}$ $\sim$ 200 M$_{\odot}$/pc$^2$, the structures of 0.15 - 0.6 pc size scales are resolved with typical resolved size of 0.3 pc. Stars form in small ($<$ 0.05 pc), dense cores, and those structures themselves could exhibit some kind of property threshold for star formation that neither the $Herschel$ data nor the stellar density estimation would have sufficient resolution to detect in most of the clouds considered here.  Of course, in the case of Mon R2 (see Figure \ref{fig:stargascore}), such a threshold that still allowed the unbroken power law of $\sim$2 to emerge from the stellar data is challenging to ponder.  Regardless, future large-scale core surveys such as the TolTEC Clouds to Cores Legacy Survey should enable this issue to be explored at a similar statistical scale to the present study, yielding more definitive core-scale results.

\section{Summary} \label{conclusion_sg}

We investigated the dependence of the star formation rate surface density ($\Sigma_{\rm{SFR}}$) on the gas mass surface density ($\Sigma_{\rm{gas}}$) in molecular clouds averaged over parsec scales. Through the $Spitzer$ Extended Solar Neighborhood Archive (SESNA) and a matching $Herschel$ archival analysis, we compiled uniform maps of the structure of young star distributions and molecular gas for twelve nearby ($<$1.5 kpc) molecular clouds with over 42000 pc$^2$ spatial coverage.

The sample of clouds span more than an order of magnitude in mass, size, and total star formation rate, and many clouds contain sub-regions spanning a range of evolutionary stages. The variation in our sample enables a robust quantification of correlation over a wider range of physical conditions than prior work has achieved. To explore the correlation, we used the star-sampled nearest neighbor technique (similar to \citealt{Gutermuth11}) where the surface densities of the star and gas are measured over the area covering the $n^{\rm{th}}$ neighbor from each stellar source. We measure the density at positions centered on the location of protostars and include all dusty YSOs (protostars and pre-main-sequence stars with disks) in the measurement of the density.

Below, we list the key conclusions of this study:

\begin{enumerate}
    \item Star formation rate surface density varies as a power-law function of the gas mass surface density for all of our sample clouds. This corresponds to an increase in the star formation efficiency with increasing gas density. Although with a small amount of deviation in some of the clouds, the power-law index is $\sim$2. Thus, we find that $\Sigma_{\rm{SFR}}$ $\propto$ $\Sigma_{\rm{gas}}^2$.
    \item Compared to previous work based on near-IR extinction maps, the scatter in the values of the power-law indices across the clouds is reduced drastically by using $Herschel$ derived H$_2$ column density maps and using densities measured at the positions of protostars uniformly extracted from the SESNA YSO catalog.
    \item Excluding highly evolved regions based on the CII/CI ratio effectively reduces edge-on disk contamination of the protostar sample and taking into account disk evolution reveals a stronger, more consistent correlation than previous works. 
    \item We do not find a column density threshold below which star formation ceases or the gradual declining star formation rate rapidly drops off. 
    \item The power-law index of $\sim$2 for the observed star-gas surface density correlation for some clouds is consistent with the prediction from an analytic model of thermal Jeans fragmentation for a sheet-like isothermal layer of gas. Consistency is also found with the predictions of a more complex hydrodynamic simulation.
\end{enumerate}

This study provides a constraining power-law index for the scaling relations between the stellar and gas surface densities at pc scales. The relation governs the star formation laws in nearby clouds and probes the physics that gives rise to the Kennicutt-Schmidt relation at much larger (e.g. kpc) scales. We will extend this study in a forthcoming paper, where we implement gas-sampled extinction contours similar to \cite{Lada10,Heiderman10,Evans14} to study the star-gas surface density correlations in the same molecular clouds as in this paper. Together, these studies will cover all the known approaches in the literature using unprecedented observational data and provide further constraints on the underlying star formation law.

\acknowledgments

We gratefully acknowledge funding support for this work from NASA ADAP awards NNX11AD14G (RAG), NNX13AF08G (RAG), NNX15AF05G(RAG, RP), 80NSSC18K1564 (RP, STM), and NNX17AF24G (RAG, RP). We also thank RP's PhD thesis committee members, Grant Wilson, Gopal Narayanan and Guy Blaylock for their valuable feedback that helped improve the manuscript. We are grateful for the anonymous referee's valuable comments and suggestions.
This research has made use of data from the Herschel Gould Belt survey (HGBS) project (http://gouldbelt-herschel.cea.fr). The HGBS is a Herschel Key Programme jointly carried out by SPIRE Specialist Astronomy Group 3 (SAG 3), scientists of several institutes in the PACS Consortium (CEA Saclay, INAF-IFSI Rome and INAF-Arcetri, KU Leuven, MPIA Heidelberg), and scientists of the Herschel Science Center (HSC).
Two
Micron All Sky Survey is a joint project of the University
of Massachusetts and the Infrared Processing and Analysis Center/California Institute of Technology, funded by the National Aeronautics and Space Administration and the National Science Foundation. This work is based in part on observations made with the $Spitzer$ Space Telescope, which is operated by the Jet Propulsion Laboratory, California Institute of Technology under a contract 1407 with NASA. Support for the IRAC instrument was provided by NASA through contract 960541 issued by the JPL.
\vspace{5mm}
\facilities{$Herschel$ (SPIRE and PACS), 2MASS, $Spitzer$ (IRAC and MIPS)}

\software{APLpy \citep{Robitaille12}, astropy \citep{Astropy13}, Matplotlib \citep{Hunter07}, NumPy (https://doi.
org/10.1109/MCSE.2011.37),
SciPy \citep{Jones01}.
          }

\bibliographystyle{aasjournal}
\bibliography{sample62}

\appendix

\section{$Herschel$ observations} \label{app:herschel}

For the Gould Belt clouds, we obtained the column density and temperature maps from the HGBS group, references for which are provided in table \ref{tab:cloudinfo_sg}. For the clouds that do not belong to the Gould Belt Survey  (distance $>$500 pc), below we provide the OBSIDs of $Herschel$ observations that we reduced:\\

\noindent
NGC 2264: 1342205056 (Level 3), 1342205057 (Level 3)\\
S140: 1342187331 (Level 2.5), 1342187332 (Level 2.5)\\
AFGL 490: 1342226619 (Level 3), 1342226620 (Level 3)\\
Cep OB3: 1342263817 (Level 2.5), 1342263818 (Level 2.5)\\
Mon R2: 1342267715 (Level 2.5), 1342267746 (Level 2.5)\\

\noindent
Cygnus-X:\\
PACS: 1342247289, 1342247288, 1342211308, 1342196917, 1342196918, 1342211307, 1342257387, 1342257385, 1342257383, 1342257382, 1342257384, 1342257386, 1342244188, 1342244170, 1342244166, 1342244169, 1342244831, 1342244168, 1342244190, 1342244832, 1342244167, 1342244191, 1342244171\\
SPIRE: 1342247288, 1342247289, 1342196917, 1342196918, 1342211307, 1342211308, 1342257382, 1342257383, 1342257384, 1342257385, 1342257386, 1342257387, 1342244189, 1342244166, 1342244167, 1342244168, 1342244169, 1342244170, 1342244171, 1342244188, 1342244190, 1342244191, 1342244831, 1342244832\\

For Cygnus-X, all PACS observations are Level 2 processed, and all SPIRE observations are Level 2.5 processed.

\section{Cropping the simulation cube} \label{app:cropping}

We have adopted molecular gas simulations in 5$^3$ pc$^3$ cubes as explained in \S \ref{hdturbulence_sg}. Observations of the three-dimensional geometry of some nearby clouds such as Perseus and Ophiuchus suggest that they are thinner along the line-of-sight, even $<$1 pc in some regions in Perseus (see \citealt{Qian15}). If so, the simulations may be too wide along the line-of-sight and need to be cropped. To address this concern, we cropped the simulated cube at different widths along the line-of-sight to see if it has any impact on the simulated star-gas density correlations. Only the gas and sink particles that are in the cropped cube are then used for obtaining the simulated star-gas surface density measurements.

\begin{figure}
    \vspace{-0.6in}
    \centering
    \includegraphics[scale=0.47]{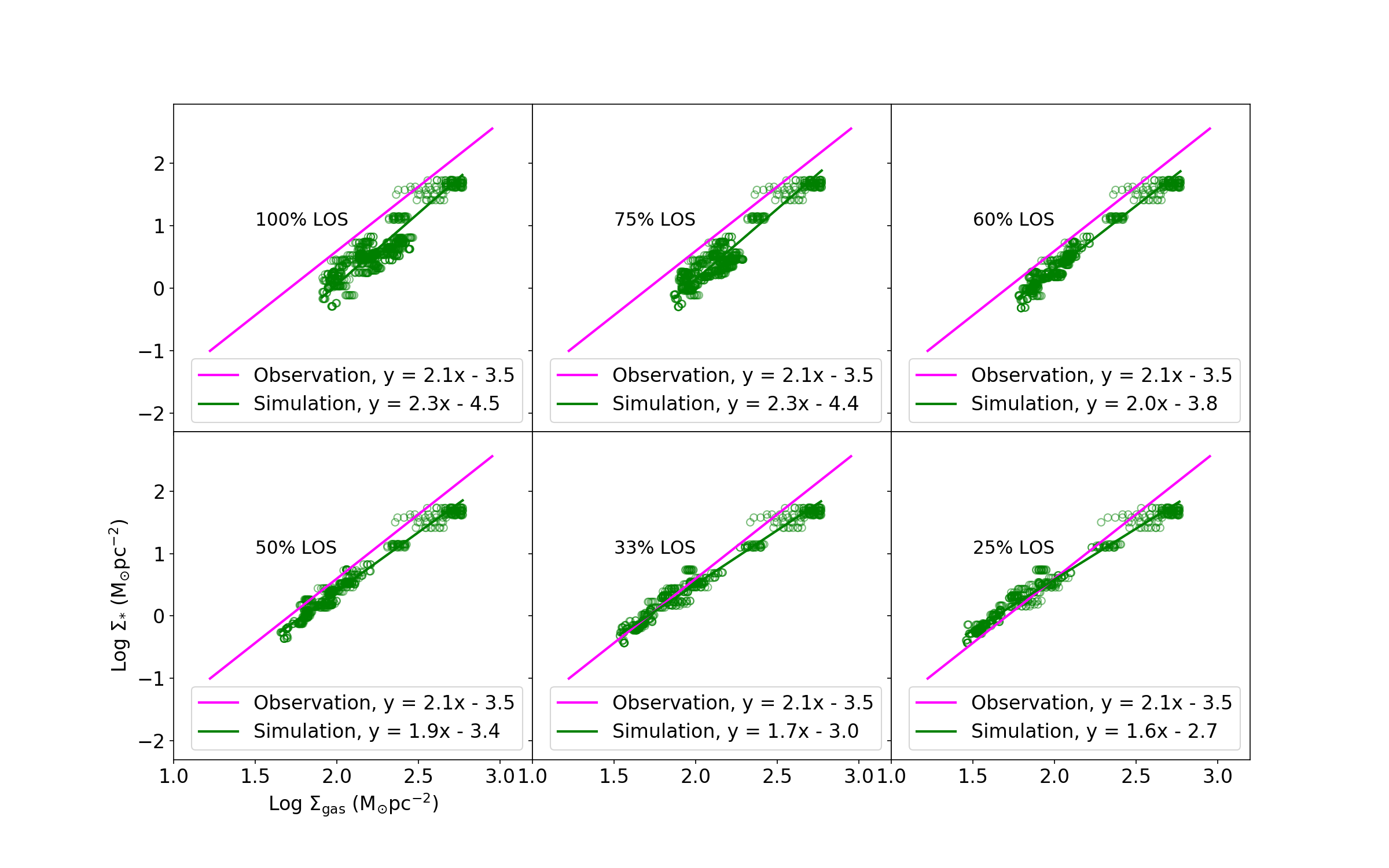}
    \vspace{-0.5in}
    \caption{Comparison of the star-gas density correlation for Mon R2 with simulation, when simulated boxes are cropped along the line-of-sight. The LOS percentage shows the percent of simulated cube used to make the plot. The magenta line represents the observed best-fit line for Mon R2, green circles represent simulated data and the green line is the best-fit line for simulated data.}
    \label{fig:stargas_sim_parsed}
\end{figure}

Figure \ref{fig:stargas_sim_parsed} shows the star-gas density correlations for a simulated cloud at a distance of 830 pc. In each panel, the line-of-sight percentage shows the fraction of the simulation cube that we kept. There is a systematic change in the slope of the best-fit line with the line-of-sight percentage. The higher column density data do not vary with the LOS percentage and only the lower column density data vary because stars form predominantly in higher density regions. The correlation becomes shallower with decreasing line-of-sight percentage. Also, as the LOS percentage decreases, lower column density data can be seen in the plot. 

Figure \ref{fig:stargas_sim_parsed} shows that we can recover the observed power-law index by cropping the simulated box at a certain percentage, for example, $\sim$60\% for the case of Mon R2. However, since cropping affects only the lower column density data and the higher column density data remains mostly intact, the simulations still do not match the observations in terms of the offset (y-intercept). Furthermore, to know the cropping portion for the box to compare with observations we need to know the geometry of the observed clouds. It should be noted that the observations do not match each other for the y-intercept either. This can also be caused by the differences in the mean cloud gas density of the clouds. Still, a factor of 2-3 could account for most of the variation, which is a relatively small difference.

\section{Cloud sample}

\begin{figure}
\centering
\includegraphics[scale=0.385]{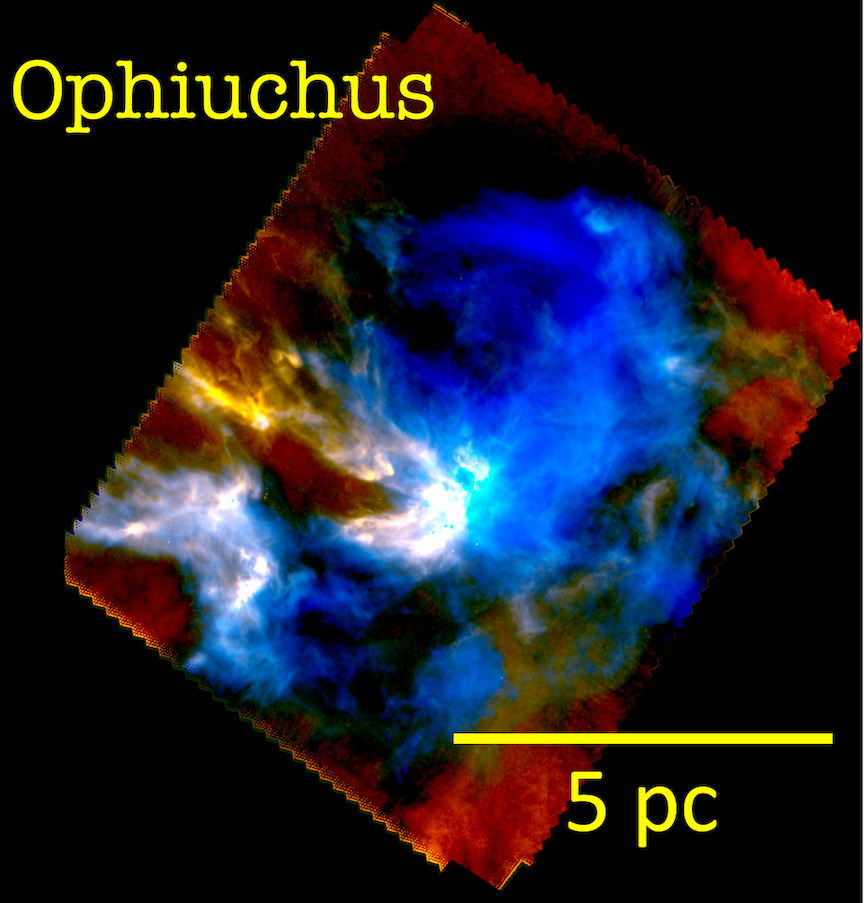}
\includegraphics[scale=0.38]{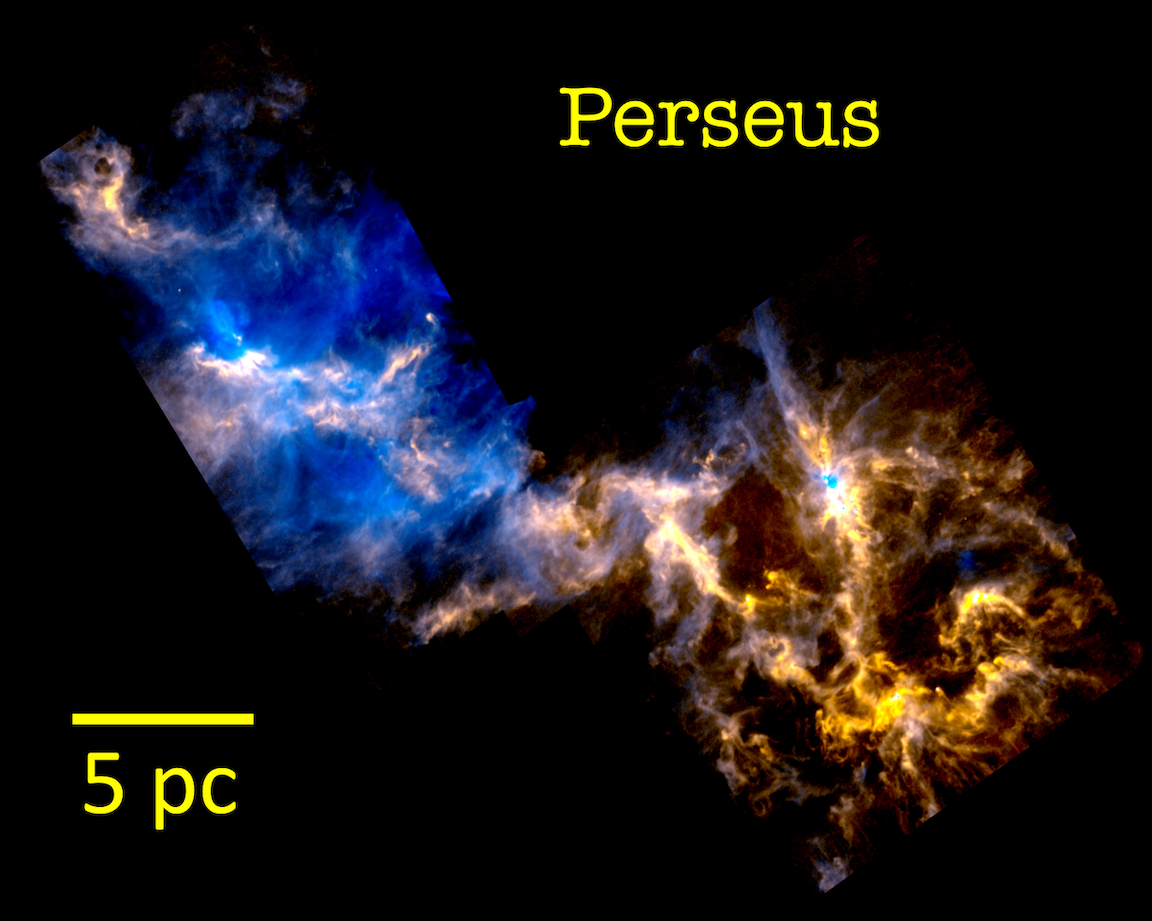}
\includegraphics[scale=0.39]{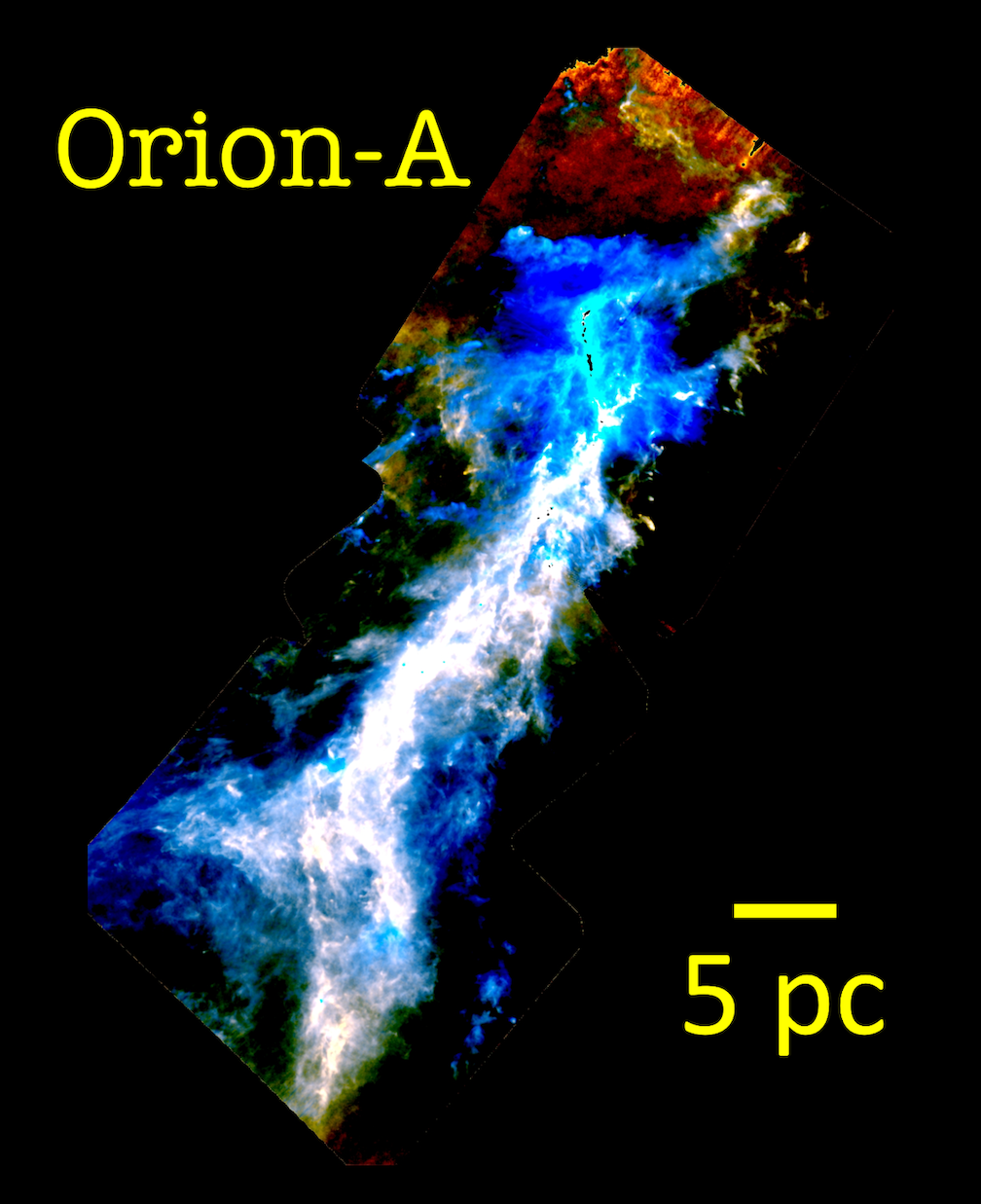}
\includegraphics[scale=0.38]{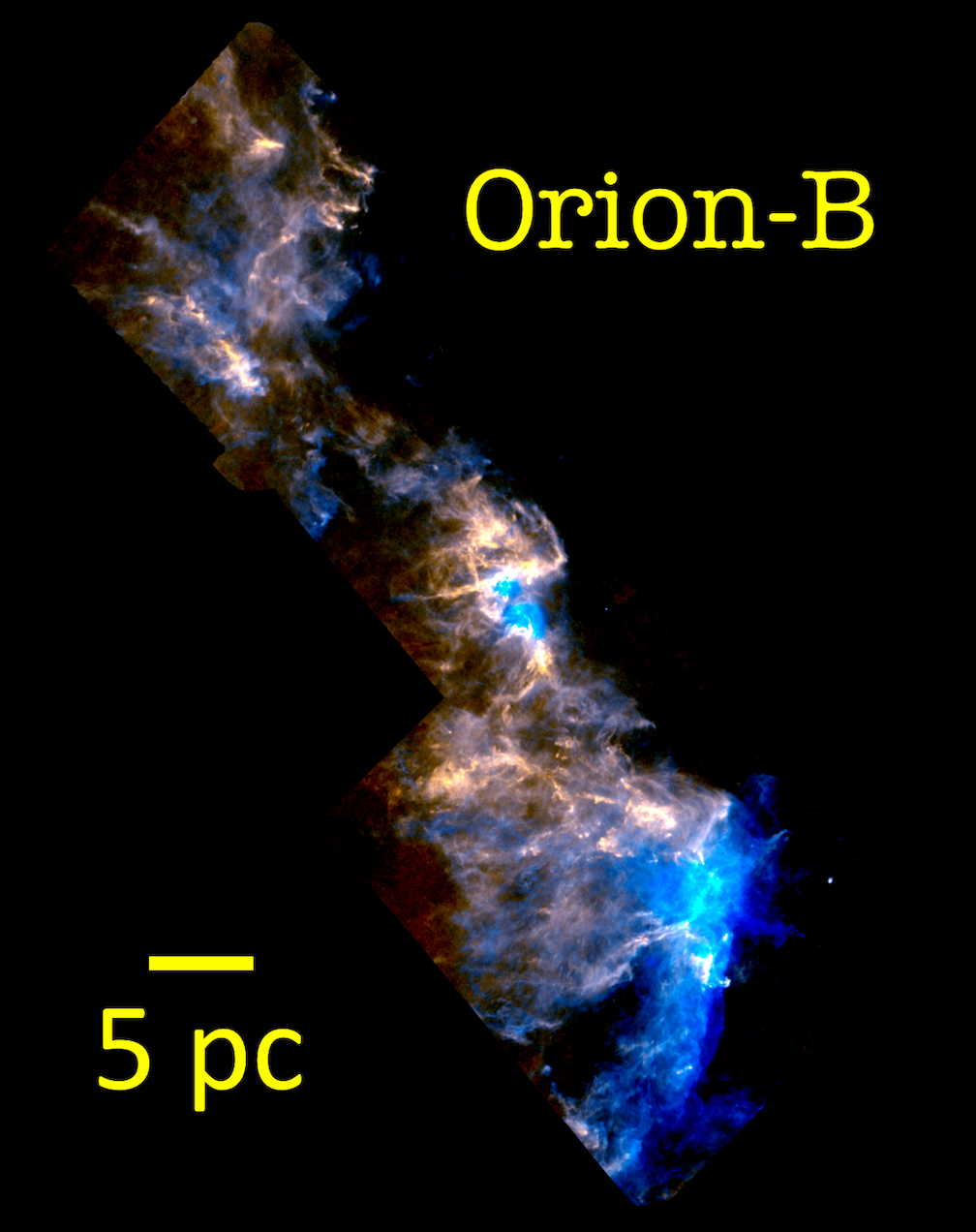}
\includegraphics[scale=0.38]{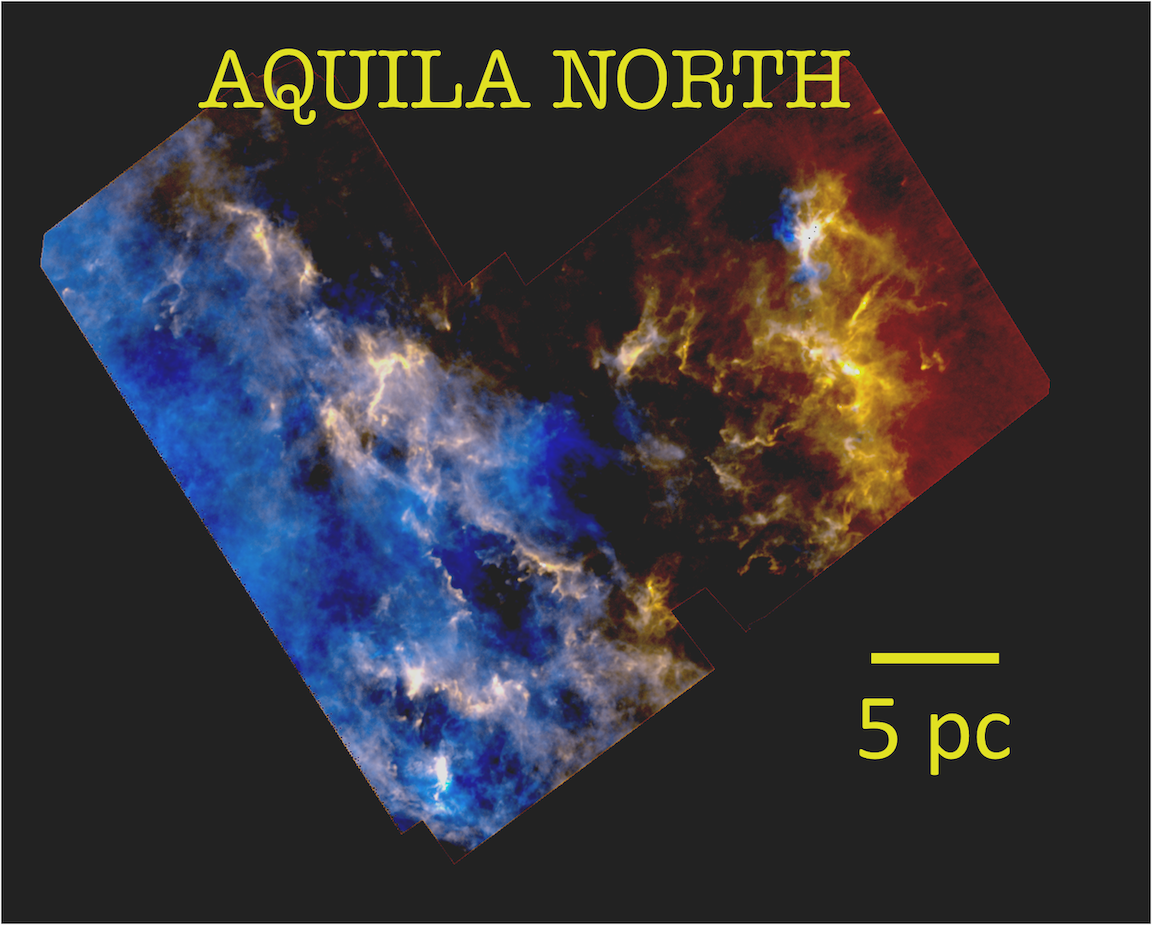}
\includegraphics[scale=0.38]{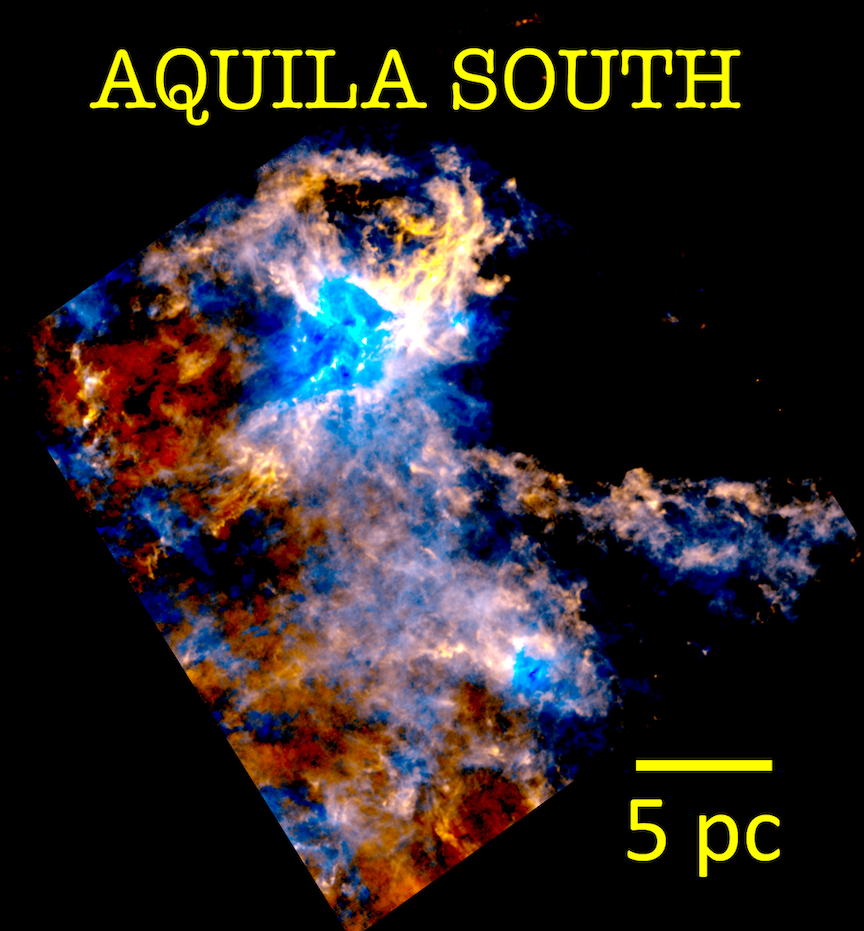}
\caption{False color images showing the column density and temperature distribution in Ophiuchus, Perseus, Orion-A, Orion-B, Aquila North and Aquila South. Column density is shown in terms of intensity of map and temperature is shown in terms of color, where $<$ 10 K regions are colored red and $>$ 20 K regions are colored blue in all the maps. The coordinates for each cloud are given in Table \ref{tab:cloudinfo_sg}.
}
\label{fig:colormap1}
\end{figure}

\begin{figure}
\centering
\includegraphics[scale=0.4]{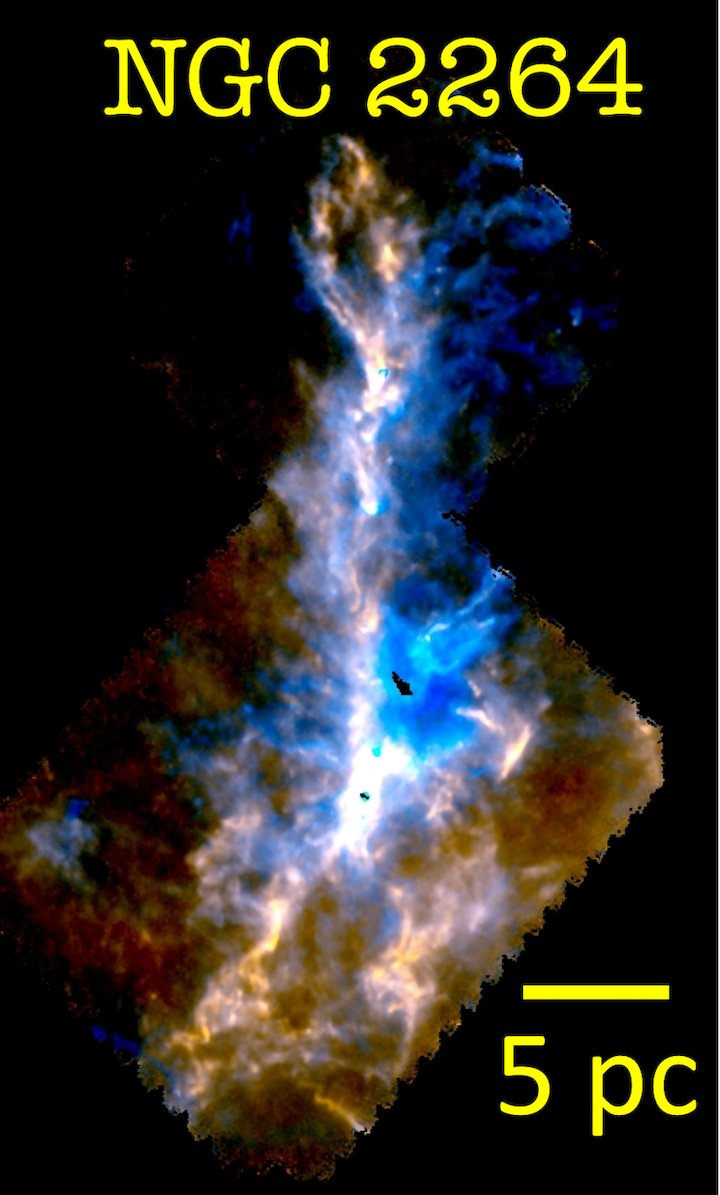}
\includegraphics[scale=0.38]{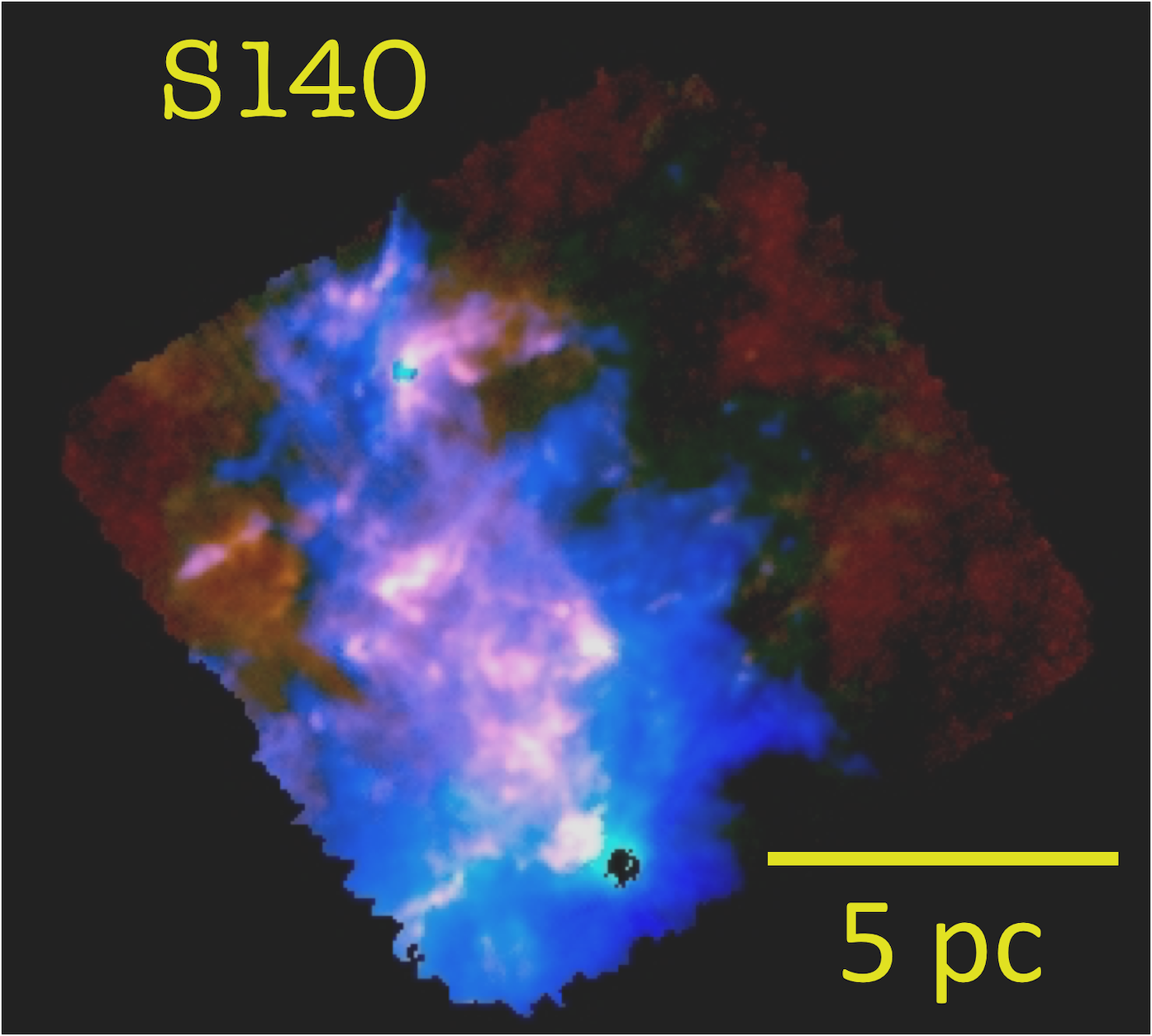}
\includegraphics[scale=0.39]{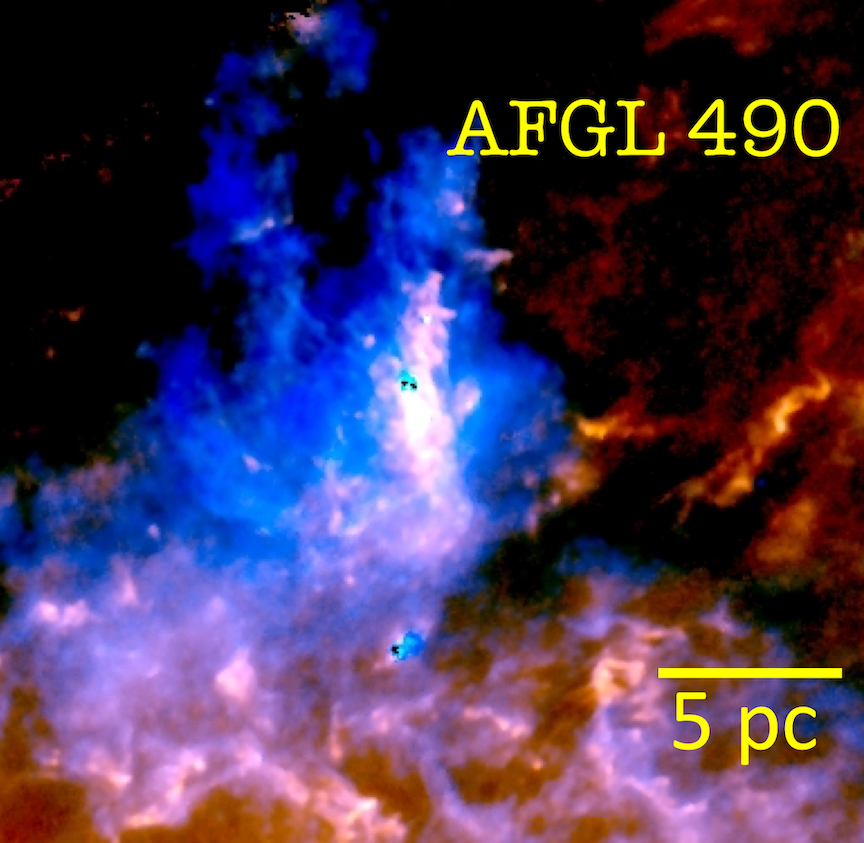}
\includegraphics[scale=0.415]{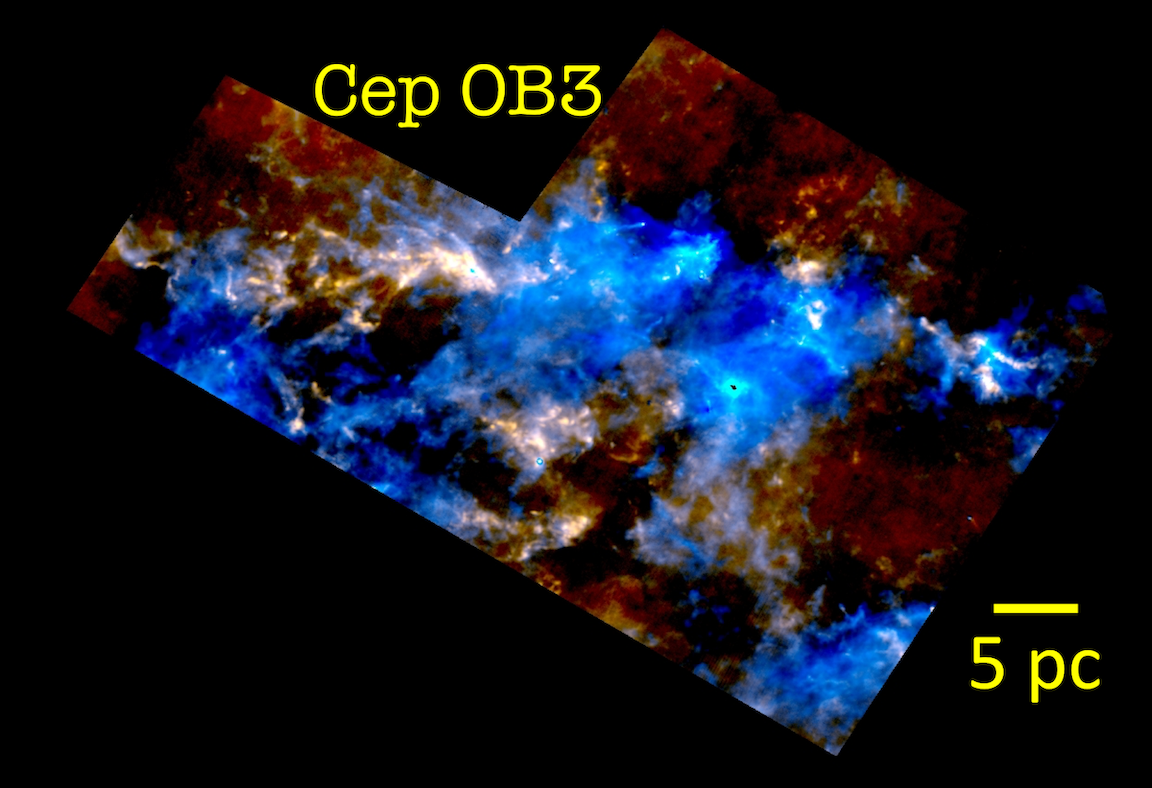}
\includegraphics[scale=0.41]{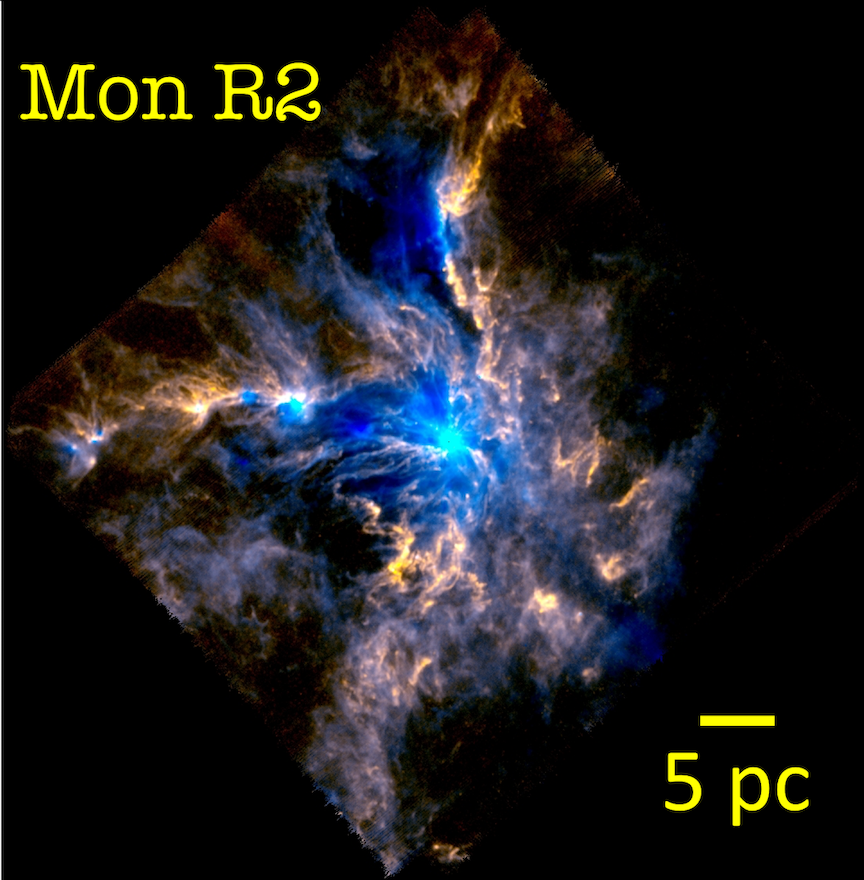}
\includegraphics[scale=0.5]{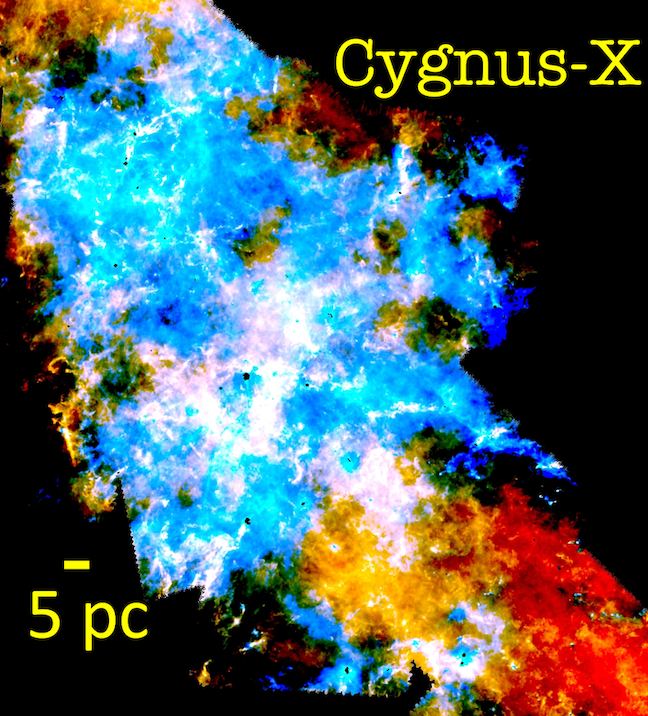}
\caption{Same as Figure \ref{fig:colormap1}, for NGC 2264, S140, AFGL 490, Cep OB3, Mon R2 and Cygnus-X. For all clouds, pixels $<$10 K are colored red. For NGC 2264, Cep OB3, and Mon R2, pixels $>$20 K are colored blue. For S140, AFGL 490 and Cygnus-X, pixels with $>$25 K temperature are colored blue. The coordinates for each cloud are given in Table \ref{tab:cloudinfo_sg}.
}
\label{fig:colormap2}
\end{figure}

Below we give brief information on the clouds that we used for this study. To visualize the density structures in the clouds, we present the combined column density and temperature map of the clouds in Figures \ref{fig:colormap1} and \ref{fig:colormap2}. In each of these images, the column density of the cloud is represented by the intensity of the image, and the temperature is represented by color. For all clouds except S140, AFGL 490 and Cygnus-X, the regions that have $<$10 K temperature are colored red and the regions that are $>$20 K are colored blue. Any other color between red and blue show the regions that are between 10 and 20 K. For S140, AFGL 490 and Cygnus-X, blue colors represent the regions that are $>$25 K.

\subsection{Ophiuchus}

Ophiuchus is the nearest molecular cloud in our sample of clouds. \cite{Ortiz18} gives an estimate of the distance to the cloud as 137.3 $\pm$ 1.2 pc using radio VLBA observations of young stars. Ophiuchus is an active, low mass star-forming cloud. The SESNA YSO catalog contains 351 YSOs in Ophiuchus, among which 70 are Class I. The cloud contains $\sim$3 $\times$ 10$^3$ M$_{\odot}$ of H$_2$ mass above 1 A$_{\rm{V}}$ and covers $\sim$140 pc$^2$ spatially, making it the smallest and least massive cloud in our sample, in addition of being the closest one.

\subsection{Perseus}

The most updated distance to Perseus is by \citet{Zucker19} using parallax measurements as a part of  $Gaia$ DR2 (d = 294 $\pm$ 17 pc). Perseus hosts low and intermediate-mass YSOs, which places it roughly between a low-mass star-forming cloud-like Taurus, and a high-mass star-forming cloud-like Orion. The cloud contains $\sim$6 $\times$ 10$^3$ M$_{\odot}$ of H$_2$ mass above 1 A$_{\rm{V}}$, which is twice of the mass of Ophiuchus. The cloud extends an area of $\sim$775 pc$^2$, and nurtures 452 YSOs according to the SESNA catalog, out of which $\sim$100 are Class I.

\subsection{Orion-A}

Orion-A is our nearest high-mass star-forming region in the Galaxy. It harbors a massive HII region popularly known as M42 inside an ``S" shaped massive dense filament, also known as the integral-shaped filament. Both the $Spitzer$ and the $Herschel$ observations cover much beyond the integral-shaped filament above  3 A$_{\rm{V}}$ contour. \citet{Yan19} used Bayesian analyses on parallax and G-band extinction measurements in  $Gaia$ DR2 to constrain the distance to Orion-A as 418 pc with 5\% systematic uncertainty. The H$_2$ column density maps are from \citet{Stutz15} which were used to show a correlation between H$_2$ column density probability density function and protostellar fraction in Orion-A. The SESNA YSO catalog lists 2394 YSOs in Orion-A, and 294 of them are Class I.


\subsection{Orion-B}

The Orion B Molecular Cloud (L 1630) is the northern one of the two major GMCs in the Orion complex. It extends over 40 $\times$ 60 pc northward from the Orion Nebula and contains several well known star-forming regions such as NGC 2071, NGC 2068, M 78 (HH 19–27), NGC 2024, and NGC 2023. The main internal heating source of L 1630 is the H II region NGC 2024 which is the second most luminous source in the Orion region and the only region in the Orion B cloud containing O stars. Orion-B lies at a similar distance as Orion-A \citep{Yan19}. $Herschel$ maps show $\sim$2 $\times$ 10$^4$ M$_{\odot}$ above 1 A$_{\rm{V}}$. The SESNA YSO catalog shows that there are 544 YSOs in Orion-B and 91 of them are Class I sources.


\subsection{Aquila North}

 Aquila North was observed as part of the Herschel Gould Belt survey (HGBS, \citealt{Andre10}) which aims at obtaining a complete census of prestellar cores and Class 0 protostars in the closest star-forming regions. Aquila North molecular cloud is a $\sim$6$^{\circ}$ $\times$ 5$^{\circ}$ star-forming complex lying close to the Galactic plane. The northern part of the cloud has a mass of $\sim$3.5 $\times$ 10$^4$ M$_{\odot}$ above 1 A$_{\rm{V}}$. Varying distances to Aquila North are reported in the literature, but we use the most updated one from \citet{Ortiz18}, $\sim$436 pc. The authors found the consistent distance by both  $Gaia$ and the VLBA for the mean parallaxes. Aquila North contains 403 YSOs in the SESNA YSO catalog, out of which 67 are protostellar.


\subsection{Aquila South}

Aquila South lies in the southern part of the Aquila Rift. It was largely unexplored until Spitzer infrared observations. Aquila South is rich in gas. $Herschel$ observations show that only $\sim$33 $\times$ 35 pc area contains more than 5 $\times$ 10$^4$ M$_{\odot}$ of gas above 1 A$_{\rm{V}}$ region. Aquila South is known to harbor two cluster-forming clumps \citep{Maury11}: Serpens South, a young protostellar cluster showing very active recent star formation and embedded in a dense filamentary cloud \citep{Gutermuth08a}, and W40 a young star cluster associated with an HII region. SESNA contains 911 YSOs in Aquila South, out of which 160 are protostellar Class I sources.


\subsection{NGC 2264}

NGC 2264 is a young cluster in the Monoceros OB1 association in Orion's Arm of our Galaxy. It contains hundreds of young stars embedded in a large molecular cloud complex presenting diffuse H$\alpha$ emission and differential interstellar extinction. The presence of Herbig–Haro objects and molecular flows confirm the active current star formation \citep{Dahm05}. Using proper motions from  $Gaia$ DR2, \citet{Kuhn19} constrained the distance to NGC 2264 to be $\sim$738 pc. The cloud contains about 2 $\times$ 10$^4$ M$_{\odot}$ of gas in $\sim$25 $\times$ 40 pc area above 1 A$_{\rm{V}}$. SESNA YSO catalog contains 558 YSOs, 100 of which are Class I.


\subsection{S140}

Sharpless 140 (S140 in short) is a relatively diffuse HII region at the edge of a much denser L1204 molecular cloud that harbors several clusters of young B stars \citep{Crampton74}.  The S140 region displays evidence of several phenomena associated with massive star formation, such as outflows and strong UV irradiation from both internal and external heating sources creating photon-dominated regions (PDRs). Using VLBI techniques \citet{Hirota08a} estimated the distance to S140 to be $\sim$ 764 pc with the help of H$_2$O masers. The $Herschel$ column density map of S140 contains $\sim$ 5 $\times$ 10$^3$ M$_{\odot}$ of gas above 1 A$_{\rm{V}}$. SESNA YSO catalog contains 531 young sources in S140, out of which 61 are protostellar.


\subsection{AFGL 490}

AFGL 490 was discovered as a bright mid-infrared (MIR) source in the AFCRL sky survey in the mid-70s and has been a target of numerous studies ever since, spanning the spectral range from optical to radio wavelengths. The region is known to show infrared CO absorption lines indicating the presence of a cooler (ca.20 K) and a warmer (ca. 100 K) gas component and P-Cygni profiles assigned to outflowing gas. The cloud extends over $\sim$ 20 $\times$ 20 pc and has a mass of $\sim$1.5 $\times$ 10$^4$ M$_{\odot}$ above 1 A$_{\rm{V}}$. The SESNA YSO catalog shows that AFGL 490 contains 319 YSOs, out of which 45 are Class I.


\subsection{Cep OB3}

Cep OB3 contains one of the nearest OB associations to our solar system enabling a good spectroscopic and kinematic study of its brightest members. \citet{Sargent79} reported sequential star formation in Cep OB3 from the proper motion survey of the region. \citet{Sargent79} subdivided the cloud into regions defined by apparently discrete peaks in the CO distribution, which she designated Cep A, B, C, D, E, and F. Our $Herschel$ maps and $Spitzer$ catalog contains these subregions and L1211, and covers $\sim$70 $\times$ 50 pc area with mass $\sim$8 $\times$ 10$^4$ M$_{\odot}$ above 1 A$_{\rm{V}}$. The SESNA YSO catalog contains 2188 YSOs in Cep OB3, out of which 205 are young Class I.


\subsection{Mon R2}

The Mon R2 region was originally identified as a group of reflection nebulae in the constellation of Monoceros. The first detailed spectroscopic and photometric study of Mon R2 nebulae was done by \citet{Racine68}, who discovered that the illuminating associated stars are mainly B-type stars, and also estimated the distance to the cloud as 830 $\pm$ 50 pc. Recent VLBI and $Gaia$ derived distance estimates agree with the former estimate within their uncertainties. $Herschel$ column density maps show the presence of $\sim$3.5 $\times$ 10$^4$ M$_{\odot}$ of H$_2$ gas above 1 A$_{\rm{V}}$. The SESNA catalog contains 931 YSOs in Mon R2, 165 of which are Class I.


\subsection{Cygnus-X}

The Cygnus X star-forming complex is a high-mass star-forming region that contains several dozen OB stars in two associations. The cloud complex is located at $\sim$1.4 kpc \citep{Rygl12} and covers an spatial extent of $\sim$140 $\times$ 160 pc. It is the closest Milky Way analog of the sorts of star-forming sites that are commonly detectable and barely spatially resolved in nearby galaxies. Thus Cygnus-X forms a bridge for studying star formation between local clouds in the Milky Way Galaxy and external galaxies. $Herschel$ column density maps show that Cygnus-X star-forming complex has a mass of $\sim$1.8 $\times$ 10$^6$ M$_{\odot}$ above 1 A$_{\rm{V}}$. Cygnus-X shows a high star formation activity with 21387 YSOs detected in the SESNA YSO catalog, out of which 2152 are Class I.


\end{document}